\documentclass[useAMS, usenatbib]{mn2e}

\usepackage[dvips]{graphicx} 
\input{epsf}

\title[Luminosity Functions in SWIRE]{Luminosity functions for galaxies and quasars in the $Spitzer$ Wide-Area Infrared Extra-galactic (SWIRE) Legacy survey}

\author[T.S.R Babbedge et al.]
      {T.S.R. Babbedge$^1$\thanks{Email: thomas.babbedge@imperial.ac.uk},
 M. Rowan-Robinson$^1$, 
 M. Vaccari$^{1,2}$,
 J. A. Surace$^3$, 
 C.J. Lonsdale$^{4,5}$,    \newauthor
 D.L. Clements$^1$,
 D. Farrah$^6$, 
 F. Fang$^3$,
 A. Franceschini$^2$,
 E. Gonzalez-Solares$^7$, \newauthor
 E. Hatziminaoglou$^8$,
 C.G. Lacey$^9$,
 S. Oliver$^{10}$,
 N. Onyett$^{10}$,
 I. P\'{e}rez--Fournon$^8$,  \newauthor
 M. Polletta$^5$,
 F. Pozzi$^{11}$,
 G. Rodighiero$^2$,
 D.L. Shupe$^3$,
 B. Siana$^4$,
 H.E. Smith$^5$\\
        $^1$Astrophysics Group, Blackett Laboratory, Imperial College London,
        Prince Consort Road, London SW7 2BW, UK.\\
         $^2$Dipartimento di Astronomia, Universita di Padova,
Vicolo Osservatorio 5, I-35122 Padua, Italy.\\
        $^3$Spitzer Science Center, MS 220--6,
California Institute of Technology, Jet Propulsion Laboratory,
Pasadena, CA 91125, USA.\\
        $^4$Infrared Processing and Analysis Center, MS 100-22,
California Institute of Technology, JPL,
Pasadena, CA 91125, USA.\\
        $^5$Center for Astrophysics \& Space Sciences, University
of California San Diego, La Jolla, CA 92093-0424, USA.\\
        $^6$Department of Astronomy, Cornell University, Space Sciences Building,
Ithaca, NY 14853, USA.\\
         $^7$Institute of Astronomy, University of Cambridge, Madingley Road, 
Cambridge, CB3 0HA, UK.\\
       $^8$Institute de Astrofisica de Canarias, C/ Via Lactea s/n, E-38200 La Laguna, Spain.\\
        $^9$Institute for Computational Cosmology, University of Durham, South Road, Durham, DH13LE, UK.\\
             $^{10}$Astronomy Centre, Department of Physics \& Astronomy,
University of Sussex, Brighton, BN1 9QH, UK.\\
      $^{11}$Dipartimento di Astronomia, Universitˆ di Bologna, viale Berti Pichat 6, I-40127 Bologna, Italy.}

   \date{Accepted 2006 May 11.  Received 2006 May 08; in original form 2006 February 16 }

\pagerange{\pageref{firstpage}--\pageref{lastpage}}
\pubyear{2006}

\begin{document}
\maketitle
\label{firstpage}
\newcommand{\mic}{\umu m}
\newcommand{\gsim}{\mathrel{\rlap{\lower4pt\hbox{\hskip1pt$\sim$}}
    \raise1pt\hbox{$>$}}}   

\begin{abstract}
We construct rest-frame luminosity functions at 3.6, 4.5, 5.8, 8 and 24$\mu$m over the redshift range 
0$<$z$<$2 for galaxies and 0$<$z$<$4 for optical QSOs, using optical and infrared data from the $Spitzer$ Wide-area InfraRed Extragalactic survey. The 3.6 and 4.5$\mu$m galaxy LFs show evidence for moderate positive luminosity evolution up to $z\sim1.5$, consistent with the passive ageing of evolved stellar populations. 
  Their comoving luminosity density was found to evolve passively, gradually increasing out to z$\sim$0.5--1 but flattening, or even declining, at higher redshift.   Conversely, the 24$\mu$m galaxy LF, which is more sensitive to 
obscured star formation and/or AGN activity, undergoes strong positive evolution, with the derived IR energy density and SFR density $\propto$ (1+z)$^\gamma$ with $\gamma$=4.5$^{+0.7}_{-0.6}$ and the majority of this evolution occurring since z$\sim$1.  Optical QSOs, however, show positive luminosity evolution in all bands, out to the highest redshifts (3$<$z$<$4).  Modelling as  L$^*\propto$(1+z)$^\gamma$ gave $\gamma$=1.3$^{+0.1}_{-0.1}$ at 3.6$\mic$, $\gamma$=1.0$^{+0.1}_{-0.1}$ at $4.5\mic$ and stronger evolution at the longer wavelengths (5.8, 8 and 24$\mic$), of $\gamma\sim3$. Comparison of the galaxy LFs to predictions from a semi-analytic
model based on CDM indicate that an IMF skewed towards higher mass
star formation in bursts compared to locally is preferred.  As a
result the currently inferred massive star formation rates in distant
sub-mm sources may require substantial downwards revision.
\end{abstract}


\newcommand{\mnras}{MNRAS}
\newcommand{\apj}{ApJ}
\newcommand{\apjl}{ApJL}
\newcommand{\apjs}{ApJS}
\newcommand{\aj}{AJ}
\newcommand{\aap}{AAP}
\newcommand{\araa}{ARA\&A}
\newcommand{\pasp}{PASP}
\newcommand{\nat}{Nature}

\def\lsim{\mathrel{\rlap{\lower4pt\hbox{\hskip1pt$\sim$}}
    \raise1pt\hbox{$<$}}}     

\section{Introduction}
\label{sec:intro}
A landmark result in astronomy over the last two decades was the launch of $COBE$ and the subsequent discovery of a strong far-IR background radiation (\citealt{Hauser1998ApJ...508...25H}; \citealt{Fixsen1998ApJ...508..123F}).  Studies of the extragalactic background suggest at least half the luminous energy generated by stars has been reprocessed into the IR by dust (e.g. \citealt{Puget1996A&A...308L...5P}, \citealt{Lagache1999A&A...344..322L}), suggesting that dust-obscured star formation was much more important at higher redshifts than locally. 
Space missions such as $IRAS$ and $ISO$ have resolved this background into large and significant populations of IR sources - to luminous IR galaxies (L$_{IR}>10^{11}$L$_{\odot}$) at z$<$1 and to ultra-luminous IR galaxies (L$_{IR}>10^{12}$L$_{\odot}$) at z$>$1 (e.g. see \citealt{Dwek1998ApJ...508..106D}; \citealt{Barger1999AJ....117..102B}; \citealt{Flores1999ApJ...517..148F}; \citealt{Chary2001ApJ...556..562C}; \citealt{Franceschini2001A&A...378....1F}; \citealt{RR2004MNRAS.345..1290R}).  These strongly evolving populations reveal that dust-shrouded activity is pivotal in understanding the formation of stars and the central black holes in galaxies.  The evolution that is required by such studies is in agreement with suggestions from longer wavelengths - the luminous sub-mm galaxies have been shown to have median redshifts $>$ 2 (\citealt{Hughes1998Natur.394..241H}; \citealt{Scott2002MNRAS.331..817S}; \citealt{Chapman2005ApJ...622..772C}), indicating that significant evolution has occurred.

Complementary results come from $Hubble$ $Space$ $Telescope$ observations, particularly when combined with $Spitzer$ observations.  These show that the bulk of 
galaxy morphologies we see locally were put into place over 0.5$<$z$<$3 (\citealt{Conselice2004ApJ...600L.139C}; \citealt{Papovich2005ApJ...631..101P}).

A combination of these results suggests that, in order to understand the mass assembly history of the 
stars and central black holes in galaxies, studies need to include IR observations and be made out to high redshift.

A basic tool in the study of galaxy populations is the construction of Luminosity Functions (LFs).  These have long been used to constrain galaxy formation models and to quantify star formation rates (SFR) and evolution (both of luminosity and number density) but have historically been focused on optical wavelengths.  Given the importance of IR-based observations to galaxy evolution we, potentially, can make large strides in understanding how galaxies assemble by constructing near, mid and far-IR galaxy LFs, and seeing how they evolve with redshift.  The shorter wavelengths tell us how evolved stellar mass builds up, whilst longer wavelengths indicate when and where the observed starbursts that are forming these stars are occurring.

To date this has been difficult since large, deep samples are required in order to trace the evolution - earlier work using $IRAS$, $ISO$ or ground-based sub-mm observations have been hampered by sensitivity, area and (for sub-mm) identification difficulties.  One overarching feature of these earlier studies has, however, been the strong evolution demonstrated by their LFs.  Some of the earliest IR LFs were constructed from the $IRAS$ satellite mission (e.g \cite{RR1987MNRAS.227..589R} derived 25, 60 and 100$\mic$ LFs).  \cite{Saunders1990MNRAS.242..318S} used $IRAS$ data to derive the 60$\mic$ and 40-120$\mic$ LFs, which were found to be indicative of strong evolution such that luminosity increases with redshift, perhaps $\propto$(1+z$)^{3\pm 1}$.  \cite{Clements2001MNRAS.325..665C} used deep 12$\mic$ $ISO$ data with follow-up optical imaging and spectroscopy to determine the mJy 12$\mic$ LF, finding that the excess (in comparison to the low-redshift $IRAS$ results) at high luminosites (higher redshift) was compatible with rapid luminosity evolution $\propto$(1+z$)^{4.5}$.  More recently,  \cite{Serjeant2004MNRAS.355..813S} used the optical-IR band-merged ELAIS Final Analysis catalogue of \cite{RR2004MNRAS.345..1290R} to calculate the ELAIS 90$\mic$ LF, finding that, for consistency with source counts, a luminosity evolution of (1+z$)^{3.4\pm 1}$ was required - consistent with the evolution in comoving volume averaged SFR at z$\lsim1$ derived from rest-frame optical and UV surveys (e.g SLOAN; \citealt{Glazebrook2003ApJ...587...55G}).

The launch of $Spitzer$ (\citealt{Werner2004ApJS..154....1W}) has given a powerful tool to astronomers.  Its wide-area deep-mapping survey capabilities means extensive progress can be made in our understanding of galaxy assembly, allowing the construction of LFs in multiple IR bands across wide luminosity and redshift ranges.  The $Spitzer$ Wide-area InfraRed Extragalactic (SWIRE: \citealt{Lonsdale2003PASP..115..897L}; \citealt{Lonsdale2004ApJS..154...54L}) survey makes use of all 7 of $Spitzer$'s imaging bands (IRAC, \citealt{Fazio2004ApJS..154...10F}: 3.6, 4.5, 5.8 and 8$\mic$.  MIPS, \citealt{Rieke2004ApJS..154...25R}: 24, 70 and 160$\mic$) and extensive ancillary data at other wavelengths across a total area of $\sim$ 49 deg$^2$ and has detected around 2.5 million infrared sources.   At the shorter IRAC wavelengths the survey can provide constraints on the more evolved stellar populations whilst at longer wavelengths it becomes increasingly sensitive to ongoing star formation.  As such, SWIRE is able to investigate the evolution of the stellar mass function, star formation rates and probe galaxy luminosity and density evolution.  With the combination of large area and high sensitivity it is uniquely able to provide good statistics for the characteristic IR sources out to z$>$1 whilst detecting rare, luminous sources out to high redshift.

  In this paper the power of SWIRE, combining both IR and optical data, is used to investigate galaxy and Type 1 active galactic nuclei (AGN) luminosity functions so that questions on their formation, evolution and relationship over a wide redshift range can be addressed.

To carry this out, the template-fitting photometric redshift method is first extended to incorporate near-IR data and then applied to band-merged 3.6 to 24$\mic$ $Spitzer$ data and associated optical photometry (U, g$\arcmin$, r$\arcmin$, i$\arcmin$ and Z) from one SWIRE field (ELAIS-N1).  The resulting photometric redshifts and best-fitting optical and IR template solutions are then used to obtain luminosity functions (LF's) and to investigate evolution with redshift.\\

In \S\ref{sec:data} the IR and optical catalogues, data calibration and band-merging are set out.  Photometric redshifts and luminosity function calculations are outlined in \S\ref{sec:templates} and \S\ref{sec:lumf}.  Luminosity functions with redshift are presented at 3.6, 4.5, 5.8, 8 and 24$\mic$ for galaxies and Type 1 AGN in \S\ref{sec:results}, and comparisons to other work are made in \S\ref{sec:lfstudies}.  The results of fitting functional forms to these LFs, evolution with redshift, energy density and star formation history are reported in \S\ref{sec:lf-fits} and overall discussions and conclusions are presented in \S\ref{sec:disc_conc}.  For this work the flat cosmological model with H$_0$=72km s$^{-1}$Mpc$^{-1}$ and $\Lambda$=0.70 is used.  Throughout this paper, the term ``AGN'' is used to refer to all Type I active galactic nuclei as determined from optical to near-IR template-fits.
\section{The data}
\label{sec:data}
\subsection{Infrared data}
\label{subsec:infrared}
In this paper we consider one field observed in the first stage of the SWIRE survey. This is 6.5 deg$^2$ of the ELAIS-N1 field (specifically, the area in common with the Isaac Newton Telescope Wide--Field Survey (INT WFS) survey, see \S\ref{subsec:optical}), centred on $16h11m+54d55m$, for which we have band-merged 3.6, 4.5, 5.8, 8 and 24$\mic$ catalogues.    Further information on the IRAC and MIPS data processing can be found in the SWIRE Data Release 2 Document (\citealt{SuraceDataDoc2005}) and in Shupe et al. (2006:in prep.).
 A plot of the IRAC, MIPS and optical coverage of ELAIS-N1 is shown in Fig. \ref{fig:lf_opIR_area}.
The depths (5 $\sigma$) are approximately 3.7, 5.4, 48 and 37.8 $\mu$Jy in the IRAC bands (3.6 to 8$\mic$) and around 230$\mu$Jy at 24$\mic$ (note this is a factor $\sim$2 higher than original estimates for 24$\mic$: \citealt{SuraceDataDoc2005}).
\subsubsection{Reliability and completeness}
\label{subsubsec:ir_relib}
\begin{center}{\it Reliability}
\end{center}
One of the most important sources of unreliability are bright star artefacts.  The majority of artefacts associated with bright stars were eliminated from the basic SWIRE source lists by tuning SExtractor to avoid their close vicinity using bespoke masks. 
 Visual inspection of random subsets of the catalogue were carried out at all wavelengths, with results indicating a reliability $>99\%$.  Investigation of the observed SEDs of SWIRE sources as compared to those of known objects was also carried out as a second test of reliability  \citep{SuraceDataDoc2005}. In order to assess the reliability of the MIPS 24$\mic$ detections, Shupe et al. (2006:in prep.) have compared the SWIRE data with deeper GTO data processed using the same methods and matching extracted sources.  Defining reliability as the ratio of all matched sources to all extracted SWIRE sources gives a reliability better than 98$\%$ for fluxes above 0.35 mJy.\\

 Based on studies with the band-merged optical-IR catalogues (e.g. \citealt{RR2005AJ....129.1183R}, hereafter RR05;  \citealt{SuraceDataDoc2005}), a $reliable$ source is here considered to be one that has a detection with S:N$>$5 in the band in question $and$ a further detection in another optical or IR band.  Although this approach will conceivably exclude real sources which are only detected in a single band this is not likely to be a significant population.  The largest effect might be expected for the 24$\mic$ population - if there is a population distinct from the shorter wavelength sources then the requirement of detection in another band will introduce a bias.  Of the 40,387 24$\mic$ sources with S:N$>$5, 772 have no detection at any other wavelength in the catalogue, or $\sim$2 per cent.  Thus any introduced bias is not severe (note that a more likely source of bias will be introduced later by requiring that the source obtains a photometric redshift).\\  

In the area in common with the optical data (see \S\ref{subsec:optical}) 331,209 $Spitzer$ sources (81\%) are defined as $reliable$.    Number counts are shown for each band in Fig. \ref{fig:num_counts}.

\begin{center}{\it Completeness}
\end{center}
The completeness of the IRAC data was evaluated by \cite{SuraceDataDoc2005} using a deeper field within ELAIS-N1 - the GOODS Validation Field (\citealt{Chary2004ApJS..154...80C}), taken as part of the Extragalactic First Look Survey (FLS) program.  
FLS and SWIRE data were compared in a 25 square arcminute area.    
The SWIRE 95$\%$ completeness level was calculated to be at 14$\mu$Jy, 15$\mu$Jy, 42$\mu$Jy, and 56$\mu$Jy for IRAC 3.6$\mic$ 4.5$\mic$ 5.8$\mic$ and 8$\mic$ respectively.  A secondary analysis via differential source counts agreed well. 
The overlap of 24$\mic$ SWIRE and FLS sources was not large enough for meaningful statisticsso in order to derive the MIPS 24$\mic$ completeness function, simulated sources were added to the central high-coverage portion of the ELAIS-N1 mosaics and SExtracted, allowing completeness versus flux to be calculated (for more details see Shupe et al. 2006: in prep).  The completeness level at 500$\mu$Jy was calculated to be $\sim$97$\%$.  Shupe et al. also carry out a direct estimate by comparing SWIRE data to deeper GTO data processed in the same manner, finding good agreement with the simulations for the drop in completeness with decreasing flux.
\subsubsection{Bandmerging}
\label{subsubsec:ir_merge}
Bandmerging of IRAC and MIPS 24$\mic$ catalogues was carried out with the $Spitzer$ Science Center's bandmerge software.  
Bandmerge reliability was estimated from cumulative distributions of positional offsets \citep{SuraceDataDoc2005}.  For $>90\%$ of merges, sources within 1.5$\arcsec$ for IRAC pairs or within 3$\arcsec$ for IRAC-MIPS-24 can be considered reliable. 
\subsection{Optical data}
\label{subsec:optical}
In this paper we use optical photometry available in the ELAIS-N1 field (U, g$\arcmin$, r$\arcmin$, i$\arcmin$ and Z) from the INT WFS.  For more information on the survey and its goals see \cite{Mcmahon2001NewAR..45...97M}.  Basic reduction, photometry and catalogue information is outlined below:

The  ELAIS-N1 (16h10m+54d30m) field is approximately 9 deg$^2$ in U, g$\arcmin$, r$\arcmin$, i$\arcmin$ and Z bands.  
  These data have been processed at Cambridge Astronomical Survey Unit's reduction pipeline (CASU) as described in \cite{Irwin2001NewAR..45..105I}.
  
The mean limiting magnitudes (Vega, 5$\sigma$ in 600 seconds) in each filter across this area are 23.40(U), 24.94(g$\arcmin$), 24.04(r$\arcmin$), 23.18(i$\arcmin$), and 21.90(Z), however the completeness falls off before this, as is demonstrated by the r$\arcmin$ magnitude distribution of sources in Fig. \ref{fig:num_counts}.
\subsubsection{Photometric calibration and merging}
\label{subsubsec:calib}
Data reduction was carried out with the CASU pipeline (further details are available in \citealt{BabbedgeThesis} and \citealt{SuraceDataDoc2005}) with an overall photometric calibration for the survey at the level of ~2 per cent.  
  In order to merge the individual optical pointings into a single optical catalogue an association radius of 0.9$\arcsec$ was used (based on the seeing, ~1.17$\arcsec$, and a nearest-neighbour separation histogram). For sources that had duplicates (from overlapping tiles), the one with the lowest photometric error was retained.  
\subsection{Optical associations}
\label{subsubsec:lfassoc}
\begin{figure}
\begin{center}
\includegraphics[height=5cm]{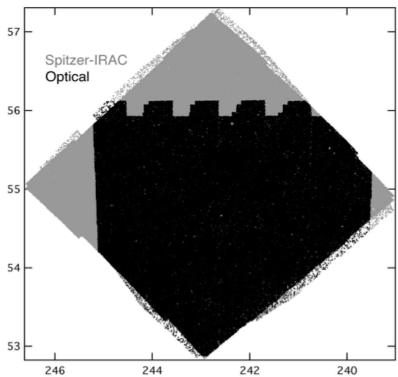}
\caption{The $Spitzer$ IRAC data (grey) and the overlap area with optical data (black).  The `crenellations' are due to the layout of the WFC CCDs.  The MIPS observations are indicated by the grey dotted region, almost entirely obscured by the IRAC area.  Image from Surace et al. 2005.
}
\label{fig:lf_opIR_area}
\end{center}
\end{figure}

The $Spitzer$ and optical observations do not have a 100$\%$ overlap.  For ELAIS-N1, the area overlap is $\sim70\%$, as illustrated in Fig. \ref{fig:lf_opIR_area}.  
The optical catalogues were associated with the band-merged SWIRE 3.6 to 24$\mic$ catalogues using the Webcmp package of IPAC's Infrared Science Archive (IRSA).  Webcmp is optimized for very fast cross-correlations, by position, between extremely large datasets.  A radius of 1.5$\arcsec$ was used for cross-identifications, a value chosen to maintain completeness whilst keeping false matches to a minimum.  Where an IR source picks up multiple optical associations, the association with the highest probability indicator value, an internal Webcmp statistic based on the match algorithm of the NASA/IPAC Extragalactic Database (NED), was retained.  This match statistic is, to first order, based on the angular separation, weighted by the catalogue uncertainties.

Completeness and reliability of these $Spitzer$-optical cross-identifications (XIDs) was investigated in \cite{SuraceDataDoc2005} 
which showed that out to the maximum association radius of 1.5$\arcsec$ the $Spitzer$-optical XIDs are essentially 100$\%$ complete, excepting the possibility of a few bright true matches at large separations.  The ratio of `false' matches to `real' matches suggest the reliability is $\sim95\%$.\\

Of the 331,209 reliable $Spitzer$ sources in the overlap area, 82,606 are optically blank (including a small fraction in INT WFS coverage gaps), meaning 248,603 (75$\%$) of the reliable $Spitzer$ sources have optical counterparts.  We then make the further requirement, in order to obtain reliable photometric redshifts in \S\ref{subsubsec:fitting}, that the source be detected in four of the bands U, g$\arcmin$, r$\arcmin$, i$\arcmin$, Z, 3.6$\mic$, 4.5$\mic$ which reduces the sample to 180,879 sources (74\%).
\subsection{Star/galaxy separation}
\label{subsec:stargal}
Star/galaxy separation is an important but complex issue, even with the luxury of multi-wavelength data.  Stellarity in a band gives a measure of the compactness of a source, however this measure is not an infallible star identifier - at faint fluxes (up to 2.5 magnitudes above the limiting magnitude) the measure begins to break down since one can then no longer reliably differentiate between extended and point-like sources.   
  Even at brighter fluxes, a star may not be classed as point-like in all bands (e.g. saturation effects) whilst conversely, though a nearby galaxy will be clearly extended, distant galaxies and quasars will also appear point-like.  Thus effective discrimination requires consideration of stellarity and saturation flags, magnitudes and colours.

For the identification of stars at fainter magnitudes it is worthwhile to note that at wavelengths covered by IRAC we are well into the Rayleigh-Jeans tail of stellar black-body curves and that stars all have essentially the same slope in their SEDs.  This means that they will be colourless in the IRAC bands (in Vega magnitudes).  As such, the stellar locus is well-defined in colour-colour space and can be used in conjunction with stellar flags to define a robust star/galaxy separator.  Here, the (3.6$\mic$/r$\arcmin$) versus (r$\arcmin$-i$\arcmin$) diagram in conjunction with stellarity is used as the basic method to identify stars and remove them from the catalogue, as detailed in RR05.  From these definitions, 41,042 of the 180,879 optical-IR sources are classed as stars (23\%), leaving 139,823 sources.
 
\begin{figure*}
\begin{center}
\includegraphics[width=15cm,height=16cm,angle=90]{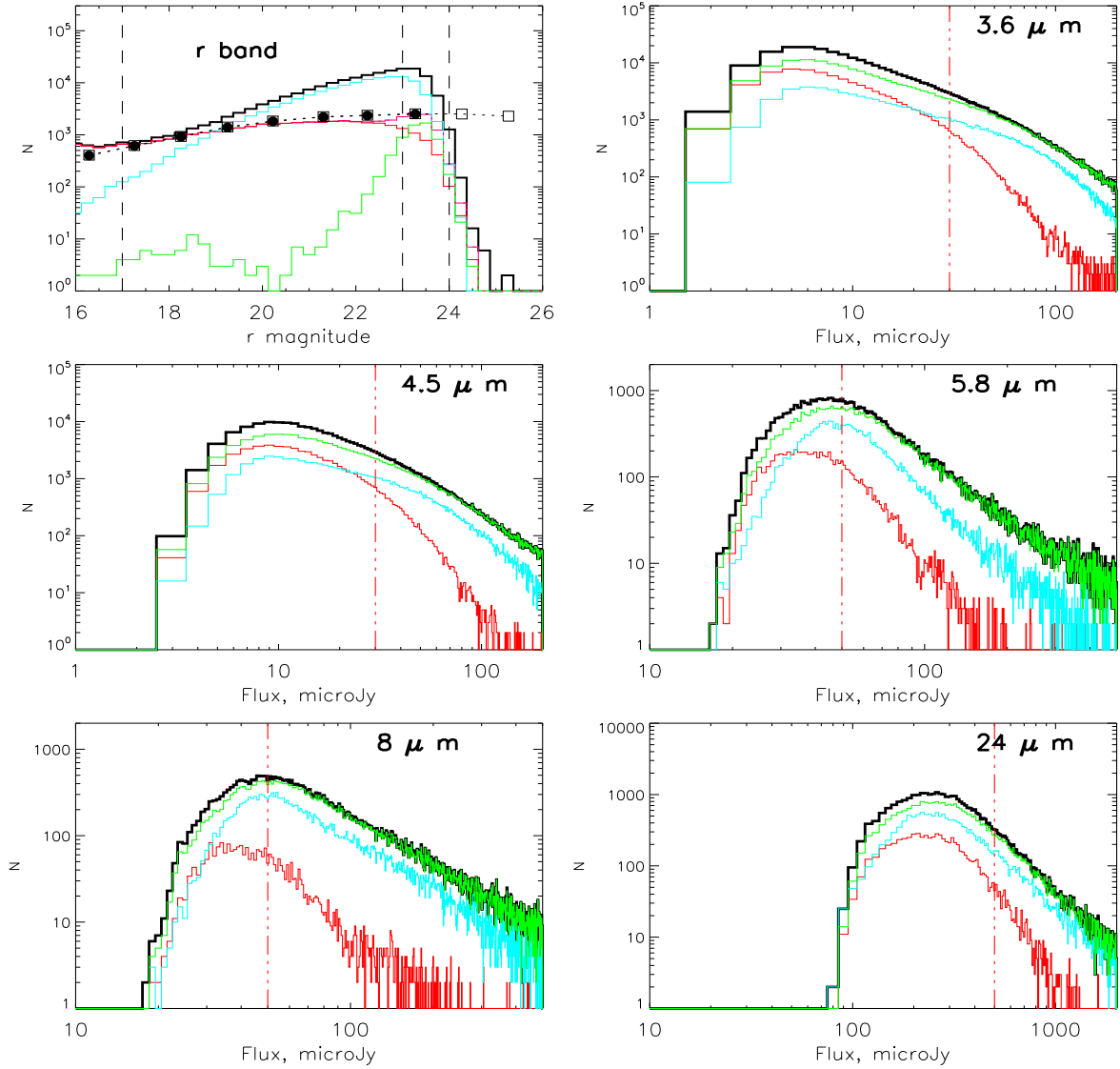}
\caption{{\bf Number count distributions.  First panel}: Magnitude distribution (Vega) of r$\arcmin$ band sources in the ELAIS-N1 full-coverage region (full optical-IRAC-MIPS overlap).  $Black$ histogram - all sources; $cyan$ - those that obtained a photometric redshift solution; $green$ - those that failed to obtain a redshift solution;  $red$ - those sources defined as stars and removed; $purple$ - the sum of stars and photometric redshift failures.  Vertical dashed lines indicate the nominal 5$\sigma$ limit of 24.04 and the magnitude cuts at r$\arcmin$=23 and 17.  The filled circles joined by a dotted line represent the stellar counts from the stellar population synthesis code of Girardi et al. (2005), empty squares are where incompleteness will have a dominant effect.  {\bf Remaining panels}:  Number counts in the full-coverage region for IRAC and MIPS 24$\mic$ sources.  $Black$ histogram - all reliable sources with a detection (S:N$>$5) in that $Spitzer$ band (will include stars); $red$ histogram - the subset that are optically blank (will include stars); $green$ histogram - the subset that have optical detections (will include stars); $cyan$ histogram - those with an r$\arcmin$ band detection of 17$<r\arcmin<23$ and which obtained a photometric redshift with $\chi^2_{red}<10$.  Vertical red dash-triple-dot lines - chosen flux below which to cut sources from the LF calculations in order to maximise completetness.  
}
\label{fig:num_counts}
\end{center}
\end{figure*}

For the optical-IR catalogue, almost all sources with r$\arcmin<17$ are stars and hence for the analyses in this  paper we assume that  all sources brighter than this are stars and removed.  In addition, all sources that are flagged as saturated in one of more optical bands are removed, since in addition to almost certainly being stars, their photometry and stellarity designations will be unreliable.  As a final cut, those sources fainter than r$\arcmin$=23 are removed in order to maintain reliability and completeness (see Fig. \ref{fig:num_counts}).  Thus we are left with a catalogue of 102,645 $extragalactic$ sources, with $17<$r$\arcmin<23$.  These sources can now be split into extended and point-like sources, where point-like sources will be either quasars or faint galaxies.  Sources are defined as point-like (and so Type 1 AGN templates are considered in addition to the galaxy SEDs in photometric redshift fitting as will be described in \S\ref{subsec:photo-z}) based on the optical class flags.  The best bands, as found in \cite{BabbedgeI} (hereafter B04), to define stellarity are g$\arcmin$, r$\arcmin$ and i$\arcmin$ so if the flag is set to `point-like' in any of these bands then the source is defined as point-like.  The goal here is simply to exclude those sources that do not look point-like in any band from having AGN templates applied since this cuts the number of spurious high-redshift AGN fits to low-redshift galaxies.
The r$\arcmin$ magnitude distribution of sources defined as stars is compared to that of the remaining sources in Fig. \ref{fig:num_counts} and it can be seen that the star distribution is relatively flat with magnitude, and dominates at the bright end.  The stellar counts from the stellar population synthesis code of \cite{Girardi2005AA...436..895G} for their CDFS field have been overlaid, converting from R band to r$\arcmin$ and scaled to the same area.  Both the CDFS and our ELAIS-N1 field are `extragalactic' fields, away from the Galactic plane and bulge, so have similar proportions of each stellar population - a normal Galactic stellar population sequence (ELAIS-N1: \citealt{Ibata2003MNRAS.340L..21I}).  We apply an additional renormalisation factor (0.74) to the total counts in the Girardi et al. model to mimick the cut placed on our multi-band catalogue prior to the identification of stars.  This cut was the requirement of detection in four of the bands U, g$\arcmin$, r$\arcmin$, i$\arcmin$, Z, 3.6$\mic$, 4.5$\mic$, which reduced our sample by a factor 0.74.  The assumption that this same factor can be applied to the Girardi et al. counts appears reasonable since the Girardi et al. counts then match our observed stellar counts at $\sim$r$\arcmin$=17, where we can be confident our stellar classification is accurate.  The open squares mark the approximate point at which incompleteness starts to have an effect at the faint end (based on the observed counts of \citealt{Groenewegen2002AA...392..741G} in the CDFS).  This empirical comparison shows that the observed star counts distribution is in agreement with that expected from an extragalactic field.  At the faint end where our ability to identify stars is less complete, some may be passed to the redshift code.  However, the majority of these appear to end up in the histogram of sources that fail to obtain a redshift - comparison of the summed distribution of stars and photometric redshift failures accounts for the majority of `missing' stars at the faint-end.  It is concluded that the contamination of stars incorrectly fit as galaxies or AGN at some redshift is small.
\section{Template fitting - a two-stage approach}
\label{sec:templates}
The emission of a galaxy at optical to near-IR and at mid-IR wavelengths is not necessarily well correlated - that is, the optical to near-IR emission will often be dominated by emission from stars, whereas at longer wavelengths re-processed emission from dust or synchrotron emission may be the dominant contribution.  Hence for a single emission mechanism the optical to mid-IR SED of a galaxy cannot be allocated to one of a set of generic templates.  Here, the analysis is separated into optical/near-IR and mid-IR, and template-fitting is carried out in a two-stage approach as in RR05:  Optical and IRAC 3.6$\mic$, 4.5$\mic$ detections are used for the photometric redshift fitting procedure (see Section \ref{subsec:photo-z}); the longer wavelengths are then used to fit IR templates to those sources defined as having an IR-excess (see Section \ref{subsec:irseds}).
\subsection{Photometric redshifts and extension to IR bands}
\label{subsec:photo-z}
In order to calculate LFs the redshifts of the sources are required.  
Here the updated version of the I{\scriptsize MP}Z code of B04 is extended to utilise some of the IR bands from $Spitzer$ and demonstrated on several sets of spectroscopic redshifts (\S\ref{subsubsec:specz_res}).  We seek to demonstrate both the validity and improvement gained by extending the template fitting photometric redshift technique to incorporate near-IR data, prior to applying it to the full ELAIS-N1 optical-$Spitzer$ catalogue.  A more in-depth study of the photometric redshift technique, its extension to IR and application to a number of large spectroscopic samples will be presented in future work (Rowan-Robinson et al. 2006 in prep.; Andreon et al. 2006 in prep.).

An important requirement for photometric redshift fitting is that the photometry in each band is measured in the same way - different aperture sizes will sample the galaxy out to different radii, and possibly in different locations.  When only using optical data, use of fixed aperture fluxes is sufficient, however if we include IRAC near-IR fluxes it is important to be comparing like with like, and these fluxes are integrated fluxes.  Hence here, for the optical bands, the aperture magnitudes corrected to the integrated magnitudes via curve-of-growth analysis are used.  This procedure could introduce errors for galaxies whose integrated SEDs differ largely from the SEDs of their central regions, but in practice the results are consistent, as found in \S\ref{subsubsec:specz_res}.
\subsubsection{Photometric redshift method}
\label{subsubsec:pzmethod}
  Improvements to the code include an iteratively calculated correction to the calibration in each band (suggested by \citealt{Ilbert2006}) and a parabolic interpolation of the redshift solution space to improve the precision, along with improvements based on studies in RR05.  One issue raised by RR05 was that for some AGN there is a significant dust torus contribution to emission at 3.6 and 4.5$\mic$ which is problematic if the code attempts to fit it as stellar emission.  One important change to the I{\scriptsize MP}Z implementation to the SWIRE data is to carry out a double pass of the code on the data, in order to deal with AGN dust tori successfully:  In the first pass, the 3.6 and 4.5$\mic$ data are not included in the fit if S(3.6)/S(r$\arcmin)>$3; in the second pass the 3.6 and 4.5$\mic$ data are included in the fit provided S(3.6)/S(r$\arcmin)<$300, except when the mid-IR fitting resulted in an AGN dust torus fit.  Mid-IR excess will instead be fit separately by more appropriate SEDs - see \S\ref{subsec:irseds}.  

\begin{figure}  
\begin{center}
\includegraphics[width=8cm]{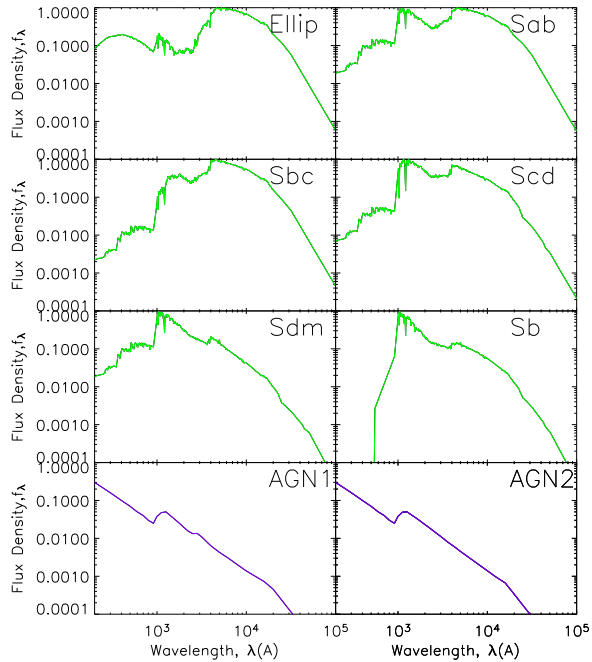}
\caption{{\bf The templates used by I{\scriptsize MP}Z}, showing the galaxy templates in green (E, Sab, Sbc, Scd, Sdm and Sb galaxies) and the two AGN templates in purple.}
\label{fig:mrr_sed_alt}
\end{center}
\end{figure}

The chosen I{\scriptsize MP}Z code setup parameters were as follows:\\
Templates:  The six galaxy templates and two (Type 1) AGN templates of B04 are shown in Fig. \ref{fig:mrr_sed_alt}.  The two AGN templates only differ from one another at $\sim3000{\rm \AA}$ where one has a `small blue bump' in its emission (blend of Balmer continuum and Fe II; \citealt{Moaz1993ApJ...404..576M}).  Determining the exact shape of the restframe far-UV spectra of AGN is problematic.  Here a simple continuation of the power-law is adopted, though at redshifts where this region enters the optical filters, the dominant effect is due to IGM absorption.  IGM treatment (\citealt{Madau1996MNRAS.283.1388M}) and Galactic extinction corrections are included by I{\scriptsize MP}Z.\\
Internal extinction:  Variable extinction is allowed in the fits via variable A$_v$ added to the templates. A$_{\rm V}$ limits of 0 to 2 in the A$_{\rm V}$ freedom are used.    As found in B04, the inclusion of A$_v$ as a parameter gives an important improvement in the accuracy of the resulting redshifts, though from \cite{Babbedge2005cnoc2} 
 it is also apparent that the derived extinction of the source is imprecise.  With the inclusion of near-IR data in the fits, A$_v$ freedom is now extended to the AGN fits, though restricted to A$_v\leq0.3$ in order to prevent aliasing with normal galaxies (it was found that otherwise low redshift galaxies could be fit as higher redshift AGN with high extinction).\\
 A$_{\rm V}$ Prior:  A prior expectation that the probability of a given value of A$_{\rm V}$ being `correct' declines as $|$A$_{\rm V}|$ moves away from 0 is applied.  This is introduced by minimising
$\chi^{2}_{red}$ + $\alpha$A$_{\rm V}^2$ rather than $\chi^2$ ($\alpha$=3 here).  This use of a prior can be viewed as a weak implementation of Bayesian methods and was reached based on degeneracy studies in B04 and \cite{RR2003MNRAS.345..819R}, hereafter RR03.\\
Magnitude limits: RR03 and B04 found it necessary to apply absolute magnitude limits to exclude unlikely solutions (such as super-luminous sources at high
redshift).  Here, the following limits are used:  Absolute magnitude limits of -16$<$M$_B$$<$[-22.5+0.3z] for galaxies and -21.7$<$M$_B$$<$-26.7 for AGN.\\
A `good' photometric redshift solution is chosen to be one with a $\chi^2_{red}<10$.  This value of $\chi^2_{red}$ is chosen as a threshold based on the results in B04 and this paper, since fits above this value are less reliable.  
\subsubsection{Comparison of spectroscopic and photometric redshifts}
\label{subsubsec:specz_res}
Investigations have been carried out for several datasets in the SWIRE fields, all of which have optical photometry, IR data from $Spitzer$ and spectroscopic redshifts.  RR05 previously demonstrated the success of the code for both an ELAIS-N1 sample and for a sample in the Lockman Validation Field (the WIYN survey; Owen et al., in prep.).  It is worth noting that the results from the WIYN comparison showed the code is successful for galaxies out to z$>$1, whilst earlier work on HDF-N (RR03) and CDF-N (B04) samples have shown success for galaxies in the range 0$<$z$<$5 and 0$<$z$<$2 respectively, and for Type 1 AGN out to z$\sim$5 (B04).

Here, updated results for several spectroscopic samples are outlined below, measuring the reliability and accuracy of the photometric redshifts via the fractional error $\Delta z/(1+z)$ for each source and examining the mean error $\overline{\Delta} z/(1+z)$ and the $rms$ scatter $\sigma_z$:
\begin{enumerate}
\item {\bf LOCKMAN VF}: In the 0.3 deg$^2$ LOCKMAN Validation Field (VF) of the SWIRE survey there are U, g$\arcmin$, r$\arcmin$ and i$\arcmin$ data, taken with the MOSAIC I camera on the Mayall 4m telescope at Kitt Peak National Observatory, and $Spitzer$ 3.6 to 24$\mu$m data.   The comparison of photometric redshifts to 269 sources with spectroscopic redshifts from the WIYN survey (Owen et al., in prep.) gives an $rms$ scatter of 0.061, with galaxies out to z$\sim$1, and Type 1 AGN out to z$\sim$3.   Compare this to typical accuracies of $\sigma_{red}\sim$0.1 for studies considering only optical photometry in the fits (RR03).
\item  
{\bf ELAIS-N1}: The spectroscopic redshift sample used in RR05 and \cite{RR2004MNRAS.345..1290R}, comprising spectroscopic redshifts in ELAIS-N1 (reported by  P\'{e}rez-Fournon et al. 2006: in prep. and Serjeant et al. 2006: in prep.; \citealt{Hatziminaoglou2005AJ....129.1198H}) with optical  U, g$\arcmin$, r$\arcmin$, i$\arcmin$, Z  and J, H, K (where available) and $Spitzer$ data.  
For this sample the total $rms$ scatter, $\sigma_{tot}$, was 0.057, where the majority of sources out to redshift z$\sim$0.5 are galaxies whilst at higher redshifts (to z$\sim$3) Type 1 AGN dominate.
\item  {\bf VVDS}: Comprises optical photometry from the CFH12K-VIRMOS survey \citep{LeFevre2004A&A...417..839L}: a deep BVRI imaging survey conducted with the CFH-12K camera in four fields.  Additionally, there are U band data from the 2.2 m telescope in La Silla and J, K data from the NTT.  The spectroscopic redshifts come from the follow-up VIRMOS-VLT Deep Survey (\citealt{LeFevre2003Msngr.111...18L}), a large spectroscopic follow-up survey which overlaps with the SWIRE data in the XMM-LSS field.  For 1394 galaxies in the range 0$<$z$<$1.5 the achieved accuracy was $\sigma_z$=0.052 with very low systematic offset.
\end{enumerate}
For all the samples, the mean systematic offset between the photometric and spectroscopic redshifts was found to be essentially zero to the precision of the photometric redshifts.  For example, the ELAIS-N1 sample has $\overline{\Delta} z/(1+z)=0.0037$.
In comparison to earlier results for the spectroscopic samples where photometric redshifts were derived from only the optical photometry (RR03; B04; \citealt{RR2004MNRAS.345..1290R}), the inclusion of the 3.6 and 4.5$\mic$ data from $Spitzer$ enables several solution degeneracies to be broken, since  data is provided on the underlying SED across a much wider observed wavelength range (0.3$\mic$ to 5$\mic$).  
The additional near-IR information has reduced the dispersion and have enabled the rejection of most of the extreme outliers resulting from optical-only results. 
This is an important improvement - even though the two IRAC bands may not have enabled a correct photometric redshift to be found for some of the original extreme outliers they have enabled the code to reject most of the original optical-only bad outliers, leaving fewer spurious redshift solutions.

From these studies it is concluded that the inclusion of the first two IRAC bands in the photometric solution reduces the $rms$ error and enables the rejection of the majority of catastrophic outliers.  
Furthermore, the inclusion of the first two IRAC bands in the solution reduces the $rms$ scatter.  Further information on photometric redshifts and their extension to IR wavelengths for the SWIRE survey can be found in \cite{BabbedgeThesis}, RR05, Rowan-Robinson et al. (2006 in prep.) and Andreon et al. (2006 in prep.).
\subsubsection{Application to the optical-$Spitzer$ catalogue}
\label{subsubsec:fitting}
The photometric redshifts of the optical-$Spitzer$ $extragalactic$ sources have been derived from their optical and IRAC 3.6, 4.5$\mic$ detections using the updated version of I{\scriptsize MP}Z with the setup set out in \S\ref{subsubsec:pzmethod}.  
The resulting photometric redshifts are expected to be accurate to $\sim0.05$ in $log_{10}(1+z)$, based on comparison to the spectroscopic sample in  \S\ref{subsubsec:specz_res} and the work of RR05, B04 and RR03. 

Due to the aliasing and photometric variability issues noted in B04 (and also \citealt{Afonso-Luis2004MNRAS.tmp..370A}) the accuracy for the AGN will be lower, $\sim0.2$ in $log_{10}(1+z)$.  The study of SWIRE/$Chandra$ sources by \cite{Franceshini2005AJ....129.2074F} (F05) also found that the photometric redshift accuracy of their AGN was low, whilst \cite{Kitsionas2005AA434475K} found that, for their X-ray selected XMM-$Newton$/2dF sample, the photometric redshift accuracy for AGN was $<0.2$ in $\Delta z/(1+z)$ for 75 per cent of the sample, with indications of the need for a new set of templates in order to model the outliers.
Here, in order to successfully deal with cases where there is significant dust torus contribution to emission at 3.6 and 4.5$\mic$, a `double pass' of the catalogue through I{\scriptsize MP}Z is carried out as described in \S\ref{subsubsec:pzmethod}. 
\begin{center}{\it Redshift results}
\end{center}
Of the 139,823 extragalactic sources input to the code 96\% of sources obtain a photometric redshift solution.  For the 102,645 sources with  $17<$r$\arcmin<23$, there are only 800 redshift failures and there are 96,177 with solutions below the $\chi^2_{red}$ maximum of 10.  It is this $17<$r$\arcmin<23$ sample with redshift solutions of $\chi^2_{red}<10$ that we will make use of in the calculation of LFs (see Table \ref{table:lf_stats} for statistics).

  From the plot comparing of the r$\arcmin$ magnitude distribution of sources that did and did not obtain a redshift (first panel of Fig. \ref{fig:num_counts}) one feature is apparent:  the magnitude distribution of those sources that failed to obtain a redshift is bimodal - there is a small distribution of bright sources (r$\arcmin<$20) which most likely failed to obtain a solution due to problems with saturated photometry and miss-associations of artefacts around bright stars and there is also a distribution of fainter sources (r$\arcmin>$21) for whom failure can be attributed to the lower accuracy of photometry at fainter fluxes, the loss of bands towards the limiting depths and also to them being faint stars which were not picked out by the original star/galaxy separation.

\begin{figure*}  
\begin{center}
\includegraphics[height=16cm,width=7cm,angle=90]{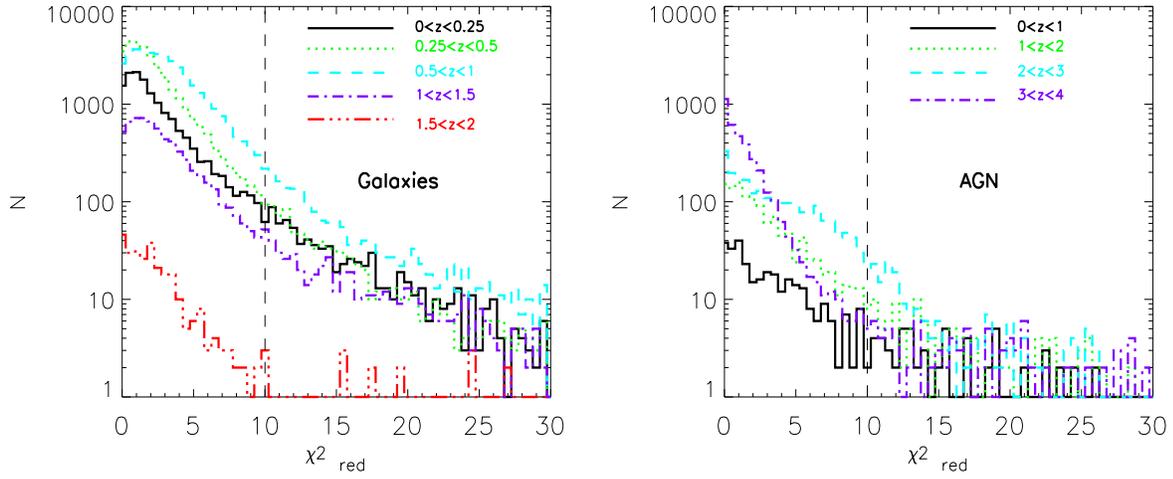}
\caption{{\bf $\chi^2_{red}$ of photometric redshifts}  for the different redshift bins considered.  {\bf Left panel} are solutions for galaxies;  redshift bins of [0 to 0.25] (black solid); [0.25 to 0.5] (green dotted); [0.5 to 1] (cyan dashed); [1 to 1.5] (purple dot-dash); [1.5 to 2] (red triple-dot-dash).   {\bf Right panel} are solutions for AGN; redshift bins are; [0 to 1] (black solid); [1 to 2] (green dotted); [2 to 3] (cyan dashed); [3 to 4] (purple dot-dashed).  Also shown is the maximum allowed $\chi^2_{red}=10$ cut-off.}
\label{fig:chi_plot}
\end{center}
\end{figure*}

We plot $\chi^2_{red}$ distributions of the photometric redshift solutions in Fig. \ref{fig:chi_plot}, split into galaxy and Type 1 AGN template solutions.  In each case, the distribution has been split into the same redshift bins as will be used for the various LF determinations in \S\ref{sec:lumf}, and the typical proportion of galaxy and AGN solutions in each redshift bin with $\chi^2_{red}<10$ is $\gsim90\%$.  
 For the galaxy solutions the slope of the $\chi^2_{red}$ distributions does not change with redshift, indicating that the applicability of the method and chosen templates remains successful out to the maximum redshift considered for the galaxy LF calculations.  The slope of the $\chi^2_{red}$ distribution for the AGN solutions does show some variation, with indications that the solutions improve at the highest redshifts considered (3 to 4).  This is attributed to the shifting into the optical bands of the Lyman break and IGM features in the AGN template which allows the AGN to be identified more readily.

\begin{figure*}
\begin{center}
\includegraphics[width=11cm,angle=90]{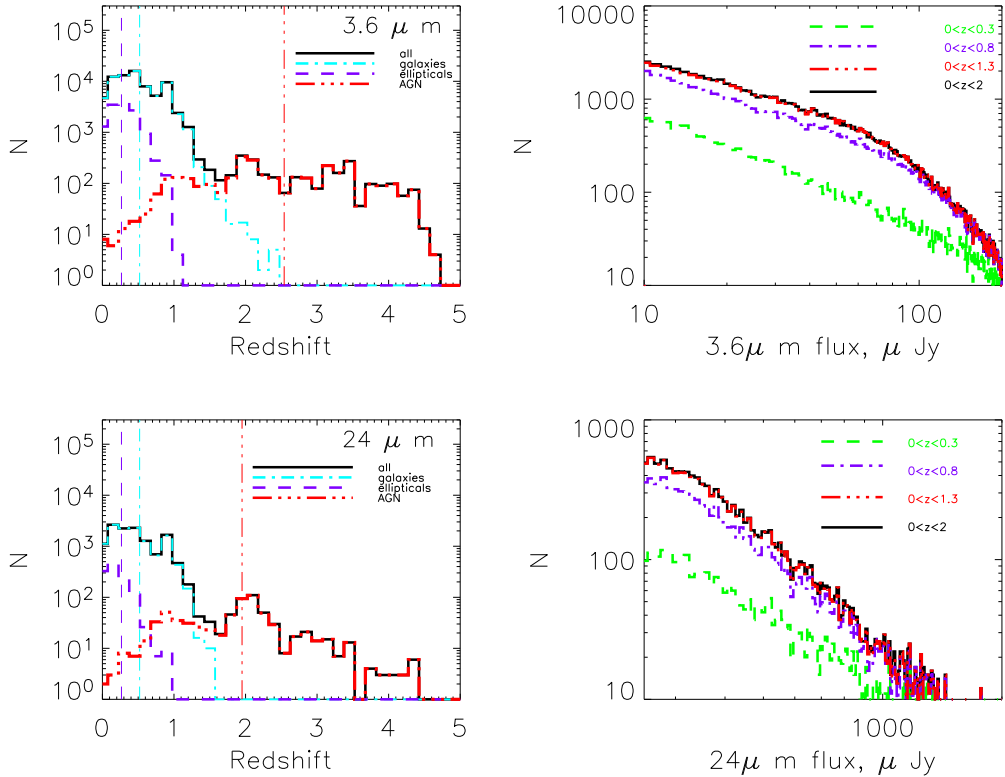}
\caption{{\bf Photometric redshift distributions in selected $Spitzer$ bands. Left panels:}  N(z) for all sources (black histogram);  Galaxy template solutions (E, Sab, Sbc, Scd, Sdm and starburst) are overplotted as light blue dot-dashed histogram; ellipticals are overplotted as purple dashed histogram; AGN solutions are overplotted as red dash-triple-dot histogram.  The mean redshift of each population is indicated by a vertical line; light-blue dot-dash line (all galaxies), purple-dashed (ellipticals), red triple-dot-dashed (AGN). {\bf Right panels:} Number counts for sources in several redshift ranges (not corrected for completeness). 
}
\label{fig:photz_dist}
\end{center}
\end{figure*}

{\it Galaxy solutions}\\
We plot redshift distributions of selected IR populations (3.6$\mic$ and 24$\mic$ sources\footnote{At points in this paper, rather than present results in all bands, we focus only on the 3.6 and 24$\mic$ results, since they represent the two ends of the populations under study.}) in the left panels of Fig. \ref{fig:photz_dist}.  The distributions for 4.5 and 5.8$\mic$ sources are similar to the 3.6$\mic$ whilst the 8$\mic$ distribution starts to move towards the distribution seen for the 24$\mic$ sources.  The galaxy populations, with $\langle$z$_{gal}\rangle$=0.55, extend out to a redshift of around 1 before dropping away.  For comparison, the 24$\micron$-selected (to 83$\mu$Jy) galaxy population of \cite{PerezGonzalez2005ApJ...630...82P} (PG05) peaked at a redshift of of 0.6--1, and decayed monotonically from redshift 1 to 3.    A comparison of the galaxy redshift distributions to the models of \cite{RR2001ApJ...549..745R}, \cite{Xu2003ApJ...587...90X} and \cite{Pozzi2004ApJ...609..122P} is given in RR05 and indicates that the models manage to broadly reproduce the observed galaxy distribution, although the \cite{Xu2003ApJ...587...90X}  model predicts the peak to be at a higher redshift, of z$\sim1$.  

The distribution of sources indicate that below a redshift of z$\sim$0.5, $\sim$22 per cent of galaxies are early-type (fit as ellipticals).  Given that our template-fitting classification is by nature statistical rather than individual this compares favourably with detailed morphological studies of K-selected galaxies with deep $HUBBLE$ Advanced Camera for Surveys (ACS) imaging (\citealt{Cassata2005MNRAS.357..903C}: 20 per cent) and the 3.6$\mic$-selected galaxy population in the ACS field of UGC10214 (\citealt{Hatz2005MNRAS.364...47H}: 24 per cent).  In the IRAC channels, the early-type population extend out to a redshift of around unity.  Above this redshift the number of sources drops away as the sensitivity of the survey to these sources is reached. The remainder of the galaxies are spirals and starbursts.  These extend out to redshifts of around 1.5 before they start to drop below the sensitivities of the optical survey.

The proportion of early-types in the 24$\mic$ population drops to $\sim$ 15 per cent of the galaxy population.  Traditionally, the IR emission from early-types is expected to be low since they are old stellar populations with little dust or ongoing star formation, leading to some contribution to the IRAC bands but negligible contribution to 24$\mic$ (e.g IRAS observations of the Shapley-Ames sample found very little emission at 60$\mic$ for ellipticals; \citealt{deJong1984ApJ...278L..67D}; \citealt{VanDenBergh2004AJ....128.1138V}).  However, IR-luminous ellipticals are not unknown: for example \cite{Leeuw2004ApJ...612..837L}'s analysis of FIR-bright ellipticals suggested they contained cold dust in their central regions whilst \cite{Krause2003AA...402L...1K}'s observations of one of the brightest 170$\mic$ sources in the ISOPHOT Serendipity Survey showed it to be an elliptical with a postulated opaque hidden starburst in its centre.

The 24$\mic$ population of early-type galaxies in SWIRE is therefore an interesting one- it may well turn out that the frequency of dusty early-type systems is greater than previously thought. As noted in RR05, it is likely that some of the elliptical template fits to 24$\mic$ sources will be Arp220-like objects which, in the optical, have SEDs similar to elliptical-like systems due to the heavy obscuration of young stars.  Similarly, late-type galaxies with significant dust extinction would appear, optically, to be reddened/early-type.  A recent spectroscopic study of SWIRE/SDSS galaxies at low redshift \cite{Davoodi2006} has shown a significant population of IR luminous galaxies have red optical colours that cannot be explained by optical extinction or contribution from AGN.   

 The mean redshift of the elliptical, spiral and starburst populations in each band can be found in Table \ref{table:lf_stats}.

  In the right panels of Fig. \ref{fig:photz_dist} we plot flux distributions of selected IR populations (3.6$\mic$ and 24$\mic$ sources) split into redshift ranges.  These show a strong evolution from z=0 to z$\sim$0.8 but this evolution is not seen to continue to higher redshifts (as commented by PG05 and \citealt{Chary2004ApJS..154...80C}).  These, however, are those $Spitzer$ sources with an optical counterpart; the $\sim$30 per cent of sources which are optically blank might well be expected to lie at higher redshift - such as in the bimodal redshift distribution of \cite{Chary2004ApJS..154...80C} which predicts a second IR bright population centred around redshift $\sim$2.

{\it Type 1 AGN solutions}\\
Eight per cent of SWIRE sources are identified by photometric redshift fitting as optical (Type I) AGN. F05's study of faint $Chandra$ X-ray sources in ELAIS N1 found the 24$\mic$ population was 10 to 20 per cent AGN (Type 1 and 2) with a ratio of between 1:2 and 1:3 between Type 1 and 2.  Hence $\sim5$ per cent of the 24$\mic$ sources were Type 1 AGN.  As a comparison, of the 18,200 24$\mic$ sources with a good photometric redshift solution and 17$<$r$\arcmin$$<$23, we find 914 optical Type 1 AGN fits, or 5 per cent.  Although a detailed cross-analysis of X-ray--derived AGN fractions to the SED template fitting and compact morphology approach of this paper is not conducted here, we have made a simple cross-match between the photometric redshift catalogue and the 99 X-ray sources from F05.  Of the 50 matches, 16 are classed as optical Type 1 AGN from our photometric redshift SED fitting whilst F05 class 11 of these 16 as Type 1 AGN, 4 as Type 2, and 1 as a late-type galaxy.  Thus our method is successful in identifying optical Type 1 AGN. There will, unavoidably, remain some Type 2 AGN `contamination' in our galaxy samples - for example, of the four elliptical-SED sources that obtain a match to the F05 $Chandra$ sample, the F05 designation was split into early-type (2), galaxy (1) and Type 2 AGN (1) classes.  More generally, F05 found that a significant fraction (up to 40 per cent) of IR sources classified as AGNs in the X-ray do not show any AGN evidence in the IR/opical SED   
 - consider that all of our 18,200 24$\mic$ sources have an IR excess, of which 23 per cent are fit in the mid-IR as AGN dust tori (i.e. Type 1 and 2 AGN).  This implies a Type 1 to 2 ratio of 1:4 or 5 since only 5 per cent are fit as a Type 1 AGN in the optical.  This is the canonical fraction one finds from standard X-ray background (XRB) models but these tend to be driven by Type 1 AGN work, which peak at z$\sim$2 say, whereas F05 found the ratio was reduced due to XRB contribution from Type 2 sources at moderate to low redshift.  Given this we would expect the obscured Type 2 to be 'hidden' in the galaxy LF's - the reader is cautioned to keep this caveat in mind when interpreting the results, particularly at the longer $Spitzer$ wavelengths where it is notoriously difficult to disentangle AGN/starburst contributions.

The AGN are seen to be a population that extends across the full redshift range, with a slight peak in their distribution at a redshift of around two, and a tail to high redshift.  This slight rise to redshift two, and decline thereafter, is as predicted by the new version of \cite{Xu2003ApJ...587...90X}'s models when including a new evolutionary model for dusty galaxies and the SWIRE limits.  This predicted AGN distribution was given in \cite{Hatziminaoglou2005AJ....129.1198H}.

Above a redshift of around two the AGN solutions dominate over the galaxy counts, though this is mainly due to SWIRE's ability to detect them to higher redshift.
\subsection{Mid IR SED fitting}
\label{subsec:irseds}
Mid-IR template fitting is carried out in a technique described in RR05.  A source that has been fit with an optical to near-IR galaxy template, based on its U band to 4.5$\mic$ detections as set out in \S\ref{subsec:photo-z}, now has its IR excess calculated by subtracting the galaxy model fit prediction from the 4.5 to 24$\mic$ data.

\begin{figure}  
\begin{center}
\includegraphics[height=6cm]{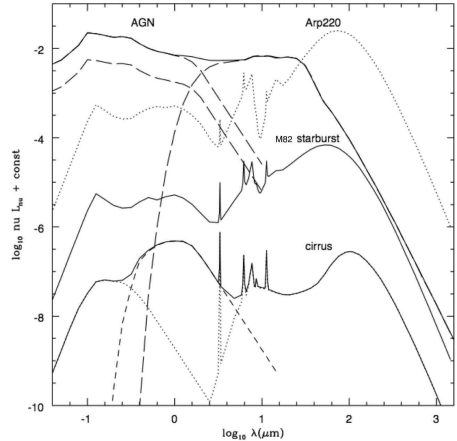}
\caption{{\bf Mid-IR spectral energy distributions:} four components - cirrus (split into low-mass [dashed line] and high-mass [dotted line] components in the optical region), M82-like starburst [solid line], Arp220-like starburst [dotted line] and AGN torus, showing assumed optical/IR ratio at log L$_{60}$=14 (upper curve) and 8.  Taken from Rowan-Robinson (2001).
}\label{fig:mir_seds}
\end{center}
\end{figure}

  At least two of these bands then need to exhibit an IR excess (one of which is required to be 8 or 24$\mic$).  
This excess is then characterised by finding the best-fitting out of a cirrus \citep{Efstathiou2003MNRAS.343..322E}, M82 and Arp220 starbursts \citep{Efstathiou2000MNRAS.313..734E} or AGN dust torus IR template \citep{RR1995MNRAS.272..737R} (shown in Fig. \ref{fig:mir_seds}).  More information on these templates can be found in RR05.  Here, it is merely noted that each template is the result of radiative transfer models and that although for $individual$ galaxies one could obtain better $individual$ fits using radiative transfer codes, we here wish to characterise only the broad types of IR population.  This `first-order' approach is in fact remarkably effective and a blind comparison of predicted 70$\mic$ flux (without making use of the 70$\mic$ information) derived from the mid-IR SED fitting method adopted here to the observed 70$\mic$ fluxes produces impressive agreement for all four mid-IR templates, across almost 2 orders of magnitude in 70$\mic$ flux (see fig. 5 of \citealt{RR2006}).

Sources are also allowed to be fit by a mixture of an M82 starburst and cirrus since it was found that, often, both components were required to properly represent the IR excess.  This mirrors the conclusions of \cite{RR1989MNRAS.238..523R} from fitting mid-IR SEDs to $IRAS$ sources, and the findings of RR05. 

For an IR SED to be fit, then, there needs to be an IR excess.  If this is not the case then an mid-IR template-based K-correction, K$_{IR}$, is not calculated for the source and the extrapolated mid-IR contribution from the optical/near-IR template fit is used instead for in the calculation of z$_{max,IR}$ (see \S\ref{sec:lumf}).  In the case of a single band IR excess, an M82 SED is assumed.  Statistics for the IR excess sources are given in Table \ref{table:lf_stats}.
\section{Luminosity function procedure}
\label{sec:lumf}
Now that the optical/near-IR and mid-IR SEDs of each source have been characterised and their photometric redshifts determined, this information can be used to calculate LFs, and to include K-corrections derived from the model templates.

Provided a selection function can be defined, a general luminosity function is calculable so long as there is no significant population of objects which remain undetected at $any$ redshift\footnote{There is also a possibility that so few galaxies of a population are detected that a selection function cannot be derived for them.  Some discussion of potential bias is given in \S\ref{subsec:bias}}.  Assuming that we have a sufficiently detailed selection function (see \S\ref{subsec:completeness}) and that no such $invisible$ population exists - a reasonable assumption given that local galaxies can all be expected to be detected in some of the multi-variate wavelengths under consideration - we can use the 1/V$_{\rm max}$ method (\citealt{Schmidt1968ApJ...151..393S} and extended by \citealt{Avni1980ApJ...235..694A}) to determine LFs for each $Spitzer$ waveband. 

 In this, the LF, $\Phi$(L), is calculated using the 1/V$_{\rm max}$ method where V$_{max}$ is the volume corresponding to the maximum redshift at which a source could be detected by the survey.  This maximum redshift may be set by either the optical limits (r$\arcmin<23$), or the IR limits, as determined by using K-corrections calculated from the optical/near-IR and mid-IR template fits.  We choose to be conservative in our choice of IR flux limits (to minimise incompleteness and maintain reliability, see \S\ref{subsec:completeness}), setting the lower IR flux limit below which a source is not considered for a given LF calculation to be:
30$\umu$Jy (3.6$\mic$); 30$\umu$Jy (4.5$\mic$); 50$\umu$Jy (5.8$\mic$); 50$\umu$Jy (8$\mic$); 500$\umu$Jy (24$\mic$).

Thus:
\begin{equation}
\label{eqn:vmax}
\Phi(L)dL=\sum_{i}\frac{1}{V_{\rm max,i}}
\end{equation}
where $\Phi$(L) is the number of objects per Mpc$^3$ in the (rest-frame) luminosity range L to (L+dL) and
\begin{equation}
\label{eqn:vi}
V_{\rm max,i}=min(V_{\rm optical,i},V_{\rm IR,i})
\end{equation}
So for the case of the rest-frame 3.6$\mic$ LF, V$_{\rm IR,i}$ is obtained by taking the source in question and redshifting it until its observed flux at 3.6$\mic$ drops below the chosen limit of 30$\mu$Jy.  V$_{optical,i}$ corresponds to the maximum redshift out to which the same source would be detectable, down to the chosen magnitude limit of r$\arcmin$=23 (in order to minimise incompleteness and photometric inaccuracy; see Fig. \ref{fig:num_counts}).

The associated $rms$ error, from Poisson statistics, is given by
\begin{equation}
\label{eqn:vi_error}
\Delta\Phi(L)=\sqrt{\sum_{i}\frac{1}{V_{\rm max,i}^2}}
\end{equation}
To find V$_{\rm optical,i}$ we find the redshift, z$_{\rm max,optical}$, at which the optical apparent magnitude, m(z), of a source of absolute magnitude M, when redshifted to a redshift, z, becomes fainter then the chosen limiting magnitude of the survey (r$\arcmin<23$), where m(z) is given by;
\begin{equation}
\label{eqn:op_mag}
m(z)=M+5log_{10}(\frac{D_L}{10pc})+2.5log_{10}K_{\rm opt}
\end{equation}
Here, D$_L$ is the luminosity distance to redshift z, and K$_{\rm opt}$ is the K-correction.
K$_{\rm opt}$ also includes the increased effect of IGM absorption at higher redshifts (\citealt{Madau1996MNRAS.283.1388M}).  K$_{\rm opt}$ is defined as:
\begin{equation}
\label{eqn:op_K}
K_{\rm opt}(z)=(1+z)\frac{\int{F(\lambda)S(\lambda)d\lambda}}{\int{F(\frac{\lambda}{1+z})S(\lambda) IGM(\lambda,z)d\lambda}}
\end{equation}
where F($\lambda$) is the optical SED template at source, S($\lambda$) is the combined filter transmission and CCD response curve, and IGM($\lambda,z$) is the redshift dependent IGM transmission.

Similarly,for V$_{\rm IR,i}$, we find the redshift, z$_{\rm max,IR}$ at which the IR flux in the IR band in question, S$_{\rm IR}$(z), of a source of absolute monochromatic luminosity, L$_{\rm IR}$, when redshifted to an arbitrary redshift, z, becomes fainter than the chosen limiting flux of the survey.  The K-correction, K$_{\rm IR}$, is defined as in Eqn. \ref{eqn:op_K}, with F($\lambda$) now the IR SED template at source and S($\lambda$) is the chosen IR filter transmission.  For IR K-corrections we do not need to incorporate IGM transmission effects for the redshifts considered.

In order to calculate the LFs correctly, however, we need to use a weighting function on the volumes in order to correct for the various selection biases that are inherent in the catalogue.  This is set out in \S\ref{subsec:completeness}.

When calculating the LFs, galaxies with redshifts above two are cut since the reliability of their photometric redshifts is lower (and the availability of spectroscopic redshifts to validate the redshift code is low) and they can have a large effect on the bright end of the LF, particularly given the low number of sources above this redshift.  This excludes $\sim0.2\%$ of the sample.  Similarly, AGN with redshifts above four are cut.  Due to photometric variability, the reduced accuracy of photometric redshift techniques when applied to AGN and the sparser opportunities for spectroscopic redshift validation this cut is prudent, excluding $\sim5\%$ of the sources fit in the optical by AGN templates.\\

We choose to split the LFs into different redshift bins in order to probe evolution.  For redshift bins the LF calculation is slightly modified, in that the minimum redshift of the redshift bin in question needs to be subtracted from the V$_{\rm max,i}$ in Eqn. \ref{eqn:vi}.  The sources are split into galaxies and AGN based on their best-fitting optical/near-IR template and for these two samples LFs are constructed in the following redshift bins:
\begin{enumerate}
\item For galaxies, LFs are constructed in the redshift bins [0 to 0.25]; [0.25 to 0.5]; [0.5 to 1]; [1 to 1.5]; [1.5 to 2].  \item For AGN (smaller number statistics and broader redshift range), LFs are constructed in the redshift bins [0 to 1]; [1 to 2]; [2 to 3]; [3 to 4].   
\end{enumerate}

\subsection{Incompleteness and selection functions}
\label{subsec:completeness}
The data set used to calculate the luminosity function in each IR band is a biased sample.  This is due to the multivariate flux limits imposed in order for a source to be included in the calculations and the requirements that it has enough information to be passed to the redshift and IR SED fitting codes $and$ obtain a solution (with $\chi^2_{red}<10$).  The general procedure for deriving an overall selection function is the same regardless of the IR band in question.  Here the 24$\mic$ LF will be used as an example:\\

The starting point for the 24$\mic$ LF calculation is the SWIRE $extragalactic$ catalogue (so stars have been removed and $17<$r$\arcmin<23$) further cut to S(24)$>$500$\umu$Jy (the chosen 24$\mic$ flux limit from \S\ref{sec:lumf}).   The probabililty that a source of a given flux at 24$\mic$ is detected by the survey, $\eta_{\rm detec}(S)$ has been determined by Shupe et al. (2006:in prep.) - for IRAC see \cite{SuraceDataDoc2005}.  Then, in order to be passed to the photometric redshift code it needs to have detections in at least four of the bands used to calculate redshift (optical, 3.6$\mic$ and 4.5$\mic$, two of which need to be 3.6$\mic$ and r$\arcmin$.  This is characterised by $\eta_{\rm pz,in}(S_{24})$.  The source then needs to obtain a solution with $\chi^2_{red<10}$, as characterised by $\eta_{\rm pz,out}(S_{24})$.  

\begin{figure}  
\begin{center}
\includegraphics{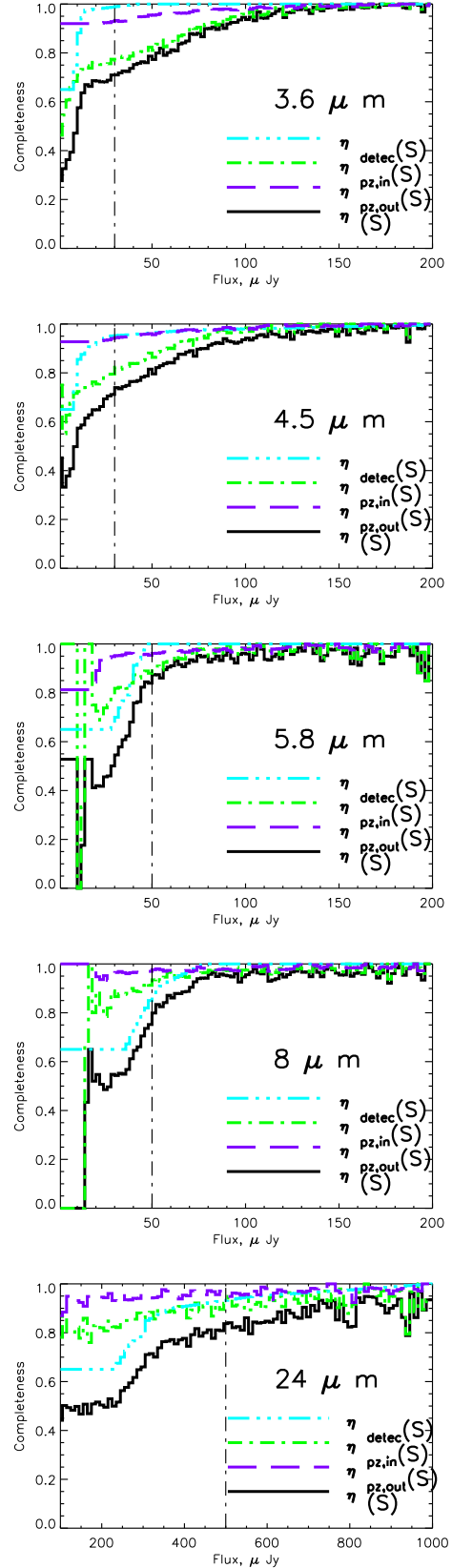}
\caption{{\bf IRAC and 24$\mic$ selection functions.} 
The chosen flux limit for each band, is shown as a vertical dot-dashed line.}
\label{fig:selec_funcs}
\end{center}
\end{figure}

Thus the probability that a given 24$\mic$ source has sufficient information to be passed to the photometric redshift code, $\eta_{\rm pz,in}(S_{24})$, is calculated in the following manner:

We take the input catalogue and construct a source count distribution, N$_{\rm all}$(S).  We then find all those sources which are eligible for photometric redshift calculation and construct a similar source count distribution, N$_{\rm pz,in}$(S).  Dividing one by the other gives the selection bias, as a function of 24$\mic$ flux: 
\begin{equation}
\label{eqn:photz_selec}
\eta_{\rm pz,in}(S)=\frac{N_{\rm pz,in}(S)}{N_{\rm all}(S)}
\end{equation}
Similarly, we can calculate the selection bias, $\eta_{\rm pz,out}(S_{24})$, introduced by requiring a source to receive a redshift solution with $\chi^2_{red}<10$, by constructing the source count distribution, N$_{\rm pz,out}$(S), of all sources which did obtain a good redshift solution, and obtain the resulting selection bias via;
\begin{equation}
\label{eqn:photzsoln_selec}
\eta_{\rm pz,out}(S)=\frac{N_{\rm pz,out}(S)}{N_{\rm pz,in}(S)}
\end{equation}
An overall selection bias, $\eta(S_{24})$, as a function of 24$\mic$ flux is then the combination of these functions, namely;
\begin{equation}
\label{eqn:overall_selec}
\eta(S)=\eta_{\rm detec}(S)\eta_{\rm pz,in}(S)\eta_{\rm pz,out}(S)
\end{equation}
The resulting selection functions for each IR band are given for IRAC and MIPS 24$\mic$ in Fig. \ref{fig:selec_funcs}.  Following \cite{Serjeant2001MNRAS.322..262S} these selection functions are used to weight the contribution of each source to the volume element integral in the LF calculation by modifying the volume elements in Eqn. \ref{eqn:vmax} to become;
\begin{equation}
\label{eqn:vmax2}
V'_{\rm max,i}=\int^{\rm z_{\rm max,i}}_{0}\eta(\rm z, S_i)\frac{\rm dV}{\rm dz}{\rm dz} 
\end{equation}

It is important to note that this process will not be able to fully account for blank-field sources - those sources detected by $Spitzer$ but absent from the optical catalogues and so having an undetermined redshift distribution.  This is one reason for the choice of conservative IR flux limits.
After implementation of the lower flux cut, requiring $17<$r$\arcmin<23$ and specifying a photometric redshift solution with $\chi^2_{red}<10$ the number of sources contributing to the LF calculation is as follows:

For 3.6$\mic$ LF calculation - 34,281 sources.

For 4.5$\mic$ LF calculation - 34,281 sources.

For 5.8$\mic$ LF calculation - 13,623 sources.

For 8$\mic$ LF calculation - 17,386 sources.

For 24$\mic$ LF calculation - 4,800 sources.
\begin{table*}  
\caption{{\bf Statistics from the catalogue}.  Note these statistics are for the full coverage region, prior to any IR flux cut imposed for LF calculations.}\label{table:lf_stats}
\scriptsize
\begin{tabular}{| l | l | l | l | l | l |}
\hline
&3.6$\mic$&4.5$\mic$&5.8$\mic$&8$\mic$&24$\mic$ \\
\hline
Enough bands for z$_{phot}$, N$_{\rm 4band}$&181,000&138,000&40,000&35,000&24,000\\
After stars removed, N$_{\rm 4band, exgal}$&140,000&105,000&26,000&25,800&23,500\\
Specify 17$<$r$\arcmin<$23, N$_{\rm 4band, exgal, r}$&103,000&81,000&23,000&24,000&20,100\\
N$_{\rm 4band, exgal}$ with z$_{phot}$ solution, N$_z$&134,000&99,500&25,600&25,500&22,800\\
N$_{\rm 4band, exgal, r}$ with z$_{phot}$ solution, N$_{z, r}$& 102,000&79,800&23,200&24,100&20,000\\
N$_{z, r}$ with $\chi_{red}^2<10$, N$_{z, r\chi}$& 96,200& 74,700& 20,500& 21,600& 18,200\\
N$_{z, r\chi}$, galaxies, N$_{gal}$ and $\langle$z$\rangle$&88,600; 0.55 &71,000;  0.55&19,700; 0.50 &20,700; 0.40 &17,300;  0.53\\
N$_{z, r\chi}$, ellipticals and $\langle$z$\rangle$& 12,900; 0.38&12,500; 0.38&6,390; 0.36&4,470; 0.30&1,630; 0.28\\
N$_{z, r\chi}$, [Sab, Sbc, Scd, Sdm] and $\langle$z$\rangle$& 52,600;  0.53&41,700; 0.54&11,100; 0.54&13,400; 0.41 &11,500; 0.52\\
N$_{z, r\chi}$, starbursts and $\langle$z$\rangle$&23,100;  0.67&16,800; 0.71&2,260;  0.66&2,850; 0.51 &4,130; 0.68\\
N$_{z, r\chi}$, optical AGN, N$_{agn}$ and $\langle$z$\rangle$&7,610; 2.78&3,680; 2.57&756; 1.99&880; 2.01&914; 2.02\\
N$_{z, r\chi}$ with IR excess, N$_{excess}$& 26,500 &26,300&14,000&20,600&18,200 \\
N$_{excess}$ fit as cirrus&4,800&4,810&3,440&4,610&4,570\\
N$_{excess}$ fit as M82&5,510&5,500&3,270&4,330&4,230\\
N$_{excess}$ fit as Arp220&2,670&2,670&1,790&2,590&2,240\\
N$_{excess}$ fit as AGN dust torus&5,160&5,150&3,400&3,470&4,290\\
N$_{excess}$ single band IR excess&8,303&8,190&2,050&5,600&2,840\\
\hline
\end{tabular}
\end{table*}
\normalsize
\subsection{Bias}
\label{subsec:bias}
LFs are the result of complex selection functions and corrections for incompleteness required to account for the biases and limits introduced by observational selection effects.

The LFs calculated here are the sum of the LFs of each galaxy type.  As such, there will be bias if certain galaxy types are not visible in a given redshift bin.  However, even when they are, there will still be a potential bias if they are observable in different absolute magnitude ranges (\citealt{Lilly1996ApJ...460L...1L}; \citealt{Brown2001AJ....122..714B}).  There are several different LF estimators, each with their own advantages and drawbacks (e.g \citealt{Willmer1997AJ....114..898W}; \citealt{Takeuchi2000ApJS..129....1T}).  The different biases each method can introduce to the LF were explored by \cite{Illbert2004MNRAS.351..541I} who considered the dependancy of limiting absolute magnitudes and the intrinsic bias introduced by the differing LFs of different galaxy populations.  
The classical 1/V$_{\rm max}$ estimator (\citealt{Schmidt1968ApJ...151..393S}), as used in this paper, accounts for the apparent magnitude limit of a survey by assuming a homogeneous galaxy distribution.  It has been shown that this assumption is not robust to clusters or voids (\citealt{Takeuchi2000ApJS..129....1T}) but this should be less of an issue for the 6.5 square degree field and large redshift range used here.  An advantage is the simplicity of the method and that it is able to give the shape and normalisation of the LF at the same time.  \cite{Illbert2004MNRAS.351..541I} found that the 1/V$_{\rm max}$ method acts to underestimate the LF at the faintest luminosity bins, biasing the global LF faint-end slope.  This was because beyond the absolute magnitude where a certain galaxy population is not observable, the method recovers the LF slope of the remaining populations.  For the 1/V$_{\rm max}$ LF calculated here, there is therefore a potential that the faint-end of the LFs has been underestimated.  In general, however, if the selection function as a function of flux and redshift is accurate, the 1/V$_{\rm max}$ will give an an unbiased estimate.\\

The second consideration for the LFs calculated here is that they are based on photometric redshifts.  One approach is to assume the photometric redshift is true for each source (as in other studies, such as \citealt{Gwyn1996ApJ...468L..77G}; \citealt{Sawicki1997AJ....113....1S}; \citealt{Mobasher1996MNRAS.282L...7M}; \citealt{Takeuchi2000ApJS..129....1T}).  This leads to potential problems with the derived LF, which have been minimised to some extent by excluding poor $\chi^2_{red}$ solutions, only calculating the LFs in the redshift range where the photometric method has been tested to a reasonable extent (z of 0-2 for galaxies and z of 0-4 for AGN), and by using generous redshift bin widths.  Despite this, the effect on the LF is a systematic one, tending to increase the number density at the bright end of the LF since the combination of redshift uncertainty and the steep slope of the LF at the bright end serves to scatter more sources to higher rather than lower luminosity.  As a result the LF becomes flatter than the intrinsic LF at the bright end (see \citealt{Chen2003ApJ...586..745C} for an in-depth discussion).
In order to obtain a more quantitative measure of the effect on the LF of the photometric redshifts we here carry out Monte Carlo simulations where each photometric redshift is assumed to be sampled from an underlying Gaussian distribution.  This will enable the effect of photometric redshift uncertainty to be quantified in the LF bins.

\begin{center}{\it Monte Carlo Analysis}
\end{center}
\begin{table}
\caption{{\bf Gaussian parameters}:  The chosen $\sigma$ of the underlying Gaussian distribution adopted for the Monte Carlo alteration of the photometric redshifts.  This $\sigma$ refers to a Gaussian centred on the photometric redshift, in log(1+z) space.}
\begin{center}
\begin{tabular}{|l |c | |l| c|}
\hline
\multicolumn{2}{c}{Galaxies} & \multicolumn{2}{c}{AGN}\\
  \hline
Redshift Interval & $\sigma$ & Redshift Interval & $\sigma$ \\
\hline
0$<$z$_{phot}<0.5$& 0.05 & 0$<$z$_{phot}<0.5$ & 0.1\\
0.5$<$z$_{phot}<1.4$& 0.1 & 0.5$<$z$_{phot}<2.5$&0.2  \\
z$_{phot}>$1.4& 0.15 & z$_{phot}>2.5$& 0.15 \\
\hline
\end{tabular}
\end{center}
\label{table:gauss}
\end{table}
Each source's photometric redshift is assumed to be sampled from an underlying Gaussian distribution with a width, $\sigma$, that is dependent on both the redshift and template type of the photometric redshift solution.  This dependence is chosen since the accuracy of the photometric redshift method varies with redshift and is poorer for AGN in comparison to galaxies (from the studies in \S\ref{subsec:photo-z} and B04).  Consider the following:
\begin{enumerate}
\item {\it Galaxies}.  At lower redshift, z$\lsim$0.5, the source will typically have good photometry and SED features such as the Balmer break are well bracketed by the filters.  At higher redshift, 0.5$<$z$<$1.4, there is still a reasonable bracketing of the main SED features by the filters, but the photometry is poorer.  Hence the accuracy of the photometric redshift is reduced, though it has been shown by numerous groups that for galaxy samples so far tested the accuracy of photometric redshift methods is within $\sigma_{\rm z}<$0.1 (\citealt{Connolly1997ApJ...486L..11C}; \citealt{FernandezSoto2001ApJS..135...41F}; RR03; B04; RR05).  At still higher redshifts, z$>$1.4, the Balmer break has left the z$\arcmin$ filter, but the Lyman limit feature will not reach the U band until a redshift of 2.5.  Coupled with increasingly faint sources, the redshift accuracy can be expected to be further reduced.
\item {\it AGN}.  Again, at low redshift, z$<$0.5, the source will be brighter and the photometry well determined (aside from photometric variability).  At higher redshift, 0.5$<$z$<$2.5, the photometric accuracy is reduced, mainly due to degeneracies between different emission lines.  At high redshift, z$>$2.5, the source has become fainter, but the Lyman limit SED feature has now entered the filters, improving the redshift identification (i.e. the Lyman dropout technique).
\end{enumerate}
Based on these considerations, the relevant Gaussian is defined for each source in Table \ref{table:gauss}.

In order to carry out the Monte Carlo analysis each source's photometric redshift is replaced by a redshift drawn randomly from the Gaussian distribution centred on the original photometric redshift (with the specification that the new redshift cannot be less than zero).  The usual LF procedure is then carried out.  This is carried out 100 times, for the LF in each $Spitzer$ band, thus producing a series of LFs for a given band.  The $rms$ spread in these LFs then represents the uncertainty introduced by using photometric redshifts instead of the true redshifts and this Monte Carlo-derived error is found to dominate over the Poisson-derived LF error, except in the least populated luminosity bins (the highest redshifts and luminosities).  
  The error in the AGN LFs will be larger since the photometric redshift technique is less accurate for AGN, due to their photometric variability over the timescale of the optical survey.
 \begin{figure}
\begin{center}
\includegraphics[angle=90]{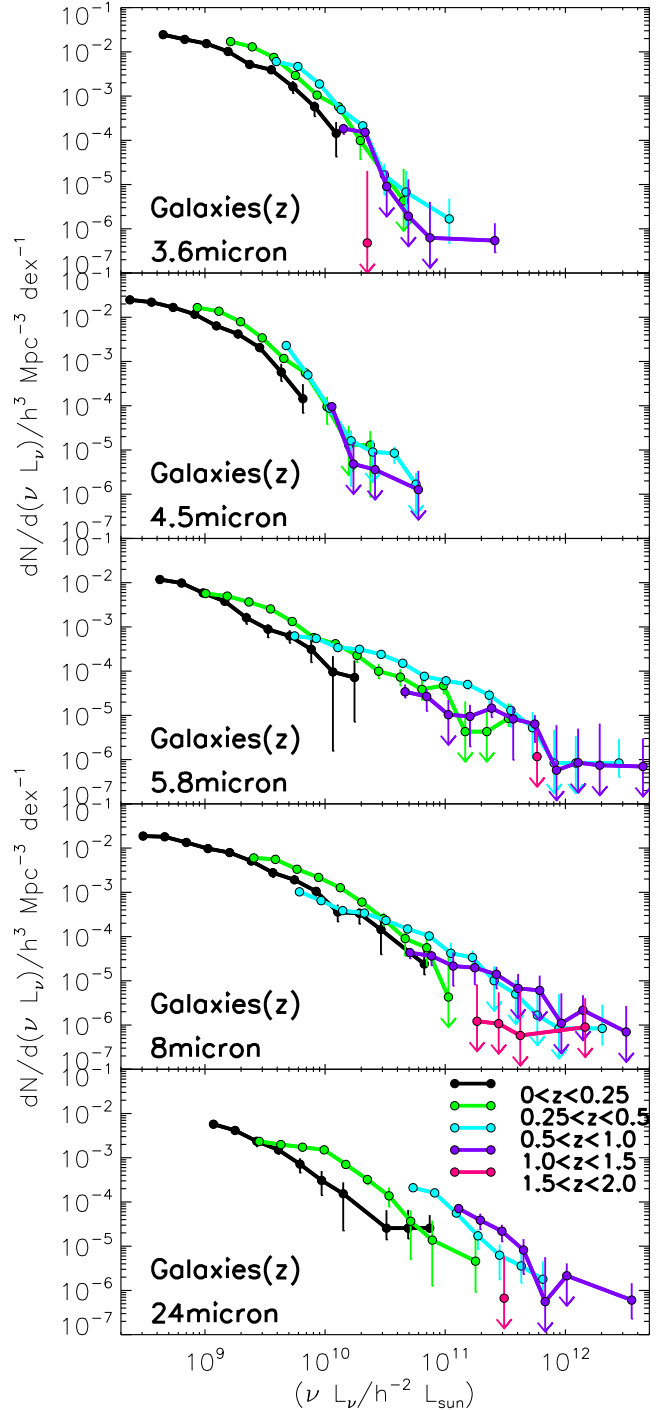}
\caption{{\bf Galaxy luminosity functions.}  Top Panel to Bottom Panel: LF for galaxies at each wavelength (excluding AGN), split into redshift bins of [0 to 0.25] (black); [0.25 to 0.5] (green); [0.5 to 1] (cyan); [1 to 1.5] (purple); [1.5 to 2] (pink).}
\label{fig:n1-galLFz}
\end{center}
\end{figure}
\section{Results}
\label{sec:results}
\subsection{Galaxy luminosity functions with redshift}
\label{subsec:lf_gal_lf}
The galaxy LFs are plotted in panels of Fig. \ref{fig:n1-galLFz}
, with errors derived from the Monte Carlo treatment.\\  
The galaxy LF at 3.6$\mic$ and 4.5$\mic$ exhibit the classical Schechter functional form \citep{Schechter1976ApJ...203..297S} which reflects the galaxy mass function (e.g. \citealt{Fontana2004AA...424...23F}; \citealt{Bundy2005ApJ...625..621B}).  That is, the 3.6 and 4.5$\mic$ LFs are sampling the stellar mass of the galaxies, since the bulk of emission at these wavelengths still arises from the older stellar populations.  At the longer $Spitzer$ wavelengths the LF is then essentially sampling the star formation rate via dust emission, and the characteristic break seen at shorter wavelengths is less prominent.  
  The bright upturn (for L$>$3E10 $\nu$L$_{\nu}$/h$^{-2}$L$_\odot$) seen in the low redshift, z$<$0.25, 24$\mic$ galaxy LF in Fig. \ref{fig:n1-galLFz} can be plausibly attributed to the contribution of obscured AGN since, as discussed in \S\ref{subsubsec:fitting}, we expect an increasing input from Type 2 AGN to our galaxy LFs at the longer $Spitzer$ wavelengths due to the classification criteria we use (optical Type 1 AGN or galaxy).

At the shorter wavelengths these LFs do not demonstrate strong evolution where the results are consistent with a passive pure luminosity evolution to higher luminosity/redshift.  The most obvious evolution is seen at the longest wavelength (24$\mic$) where for the first three redshift bins this evolution can be seen as a general strong increase in luminosity with redshift.  The two highest galaxy redshift bins (z$>$1) do not show a strong continuation of this trend.  
Across all five bands we find that the numbers of galaxies in the highest redshift bin (1.5$<$z$<$2) are not sufficient to properly characterise the galaxy LF at this epoch, but the results hint at a drop in space-density.

One might additionally expect that different types of galaxy exhibit a different luminosity function due to their differing formation and evolution history.  We can use our broad-brush SED characterisations in order examine the galaxy LFs with source type.  The photometric redshift template-fitting approach allows us to constrain the redshift, and to loosely constrain the source type - that is, the characterisations are valid in a large-scale statistical sense but have less meaning for individual sources.  The ability of photometric redshift codes to reliably recover galaxy type in addition to redshift is not well quantified, since the method often relies on clear SED features such as the Balmer break which are found across several galaxy types.  The study of \cite{Bolzonella2000A&A...363..476B} found, however, that the success rate for bursts and ellipticals was high, since they differ markedly from more intermediate types.  Thus here we choose to split the galaxy LFs by optical SED template type, into three sub-classes (named so as to avoid morphological inferences); Early-type, `Late-type' (Sab, Sbc, Scd and Sdm), and Starbursts, with this caveat in mind.

\begin{figure*}
\begin{center}
\includegraphics[height=14cm, angle=90]{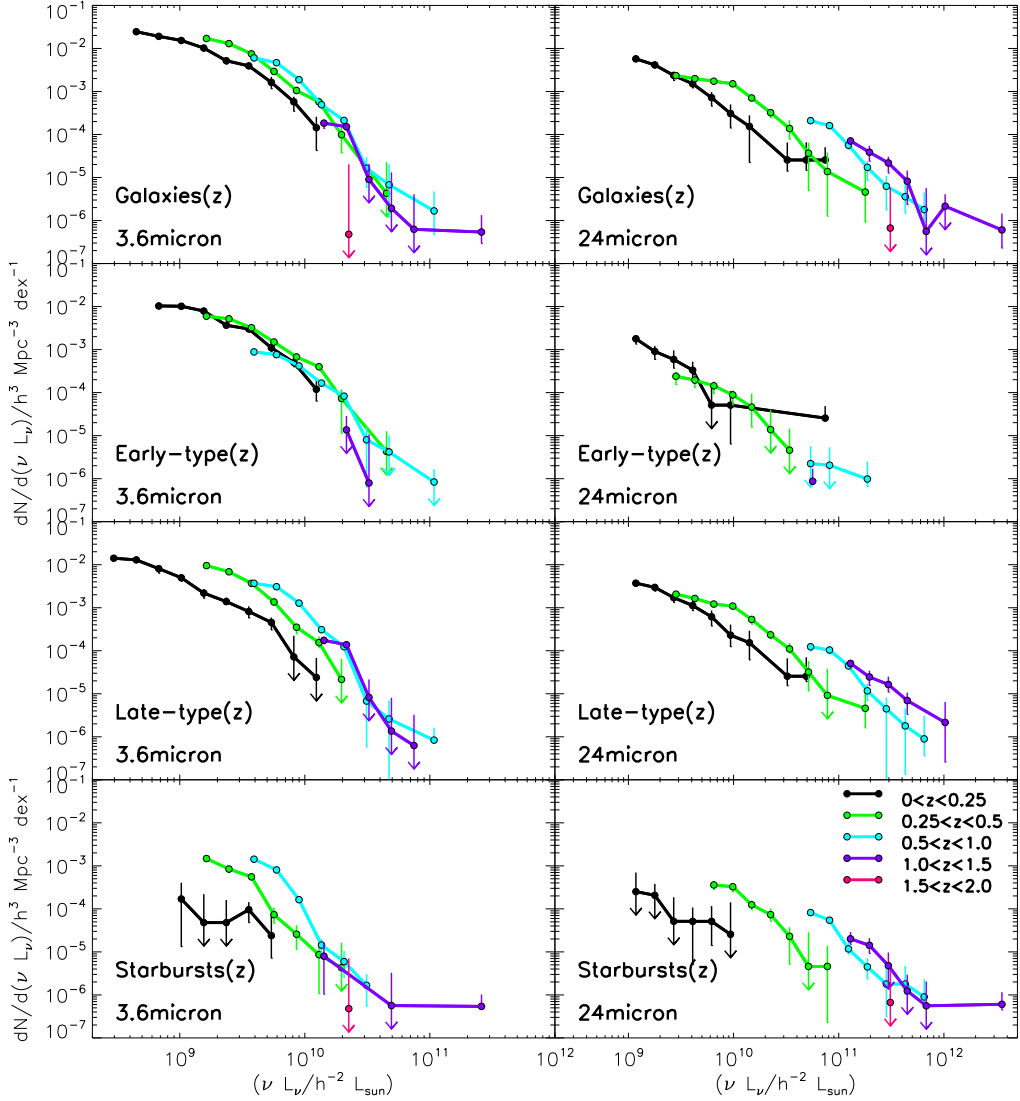}
\caption{\scriptsize{{\bf Selected galaxy luminosity functions - 3.6$\mic$ and 24$\mic$.} First row; LF for galaxies (excluding AGN), split into redshift bins of [0 to 0.25] (black); [0.25 to 0.5] (green); [0.5 to 1] (cyan); [1 to 1.5] (purple); [1.5 to 2] (pink).  Second row; LF for ellipticals, split into the same redshift bins. Third row; LF for spirals, split into the same redshift bins.  Fourth row; LF for starbursts, split into the same redshift bins.}
\label{fig:n1-galz}}
\end{center}
\end{figure*}

  The LFs of these sub-classes are plotted in sub-panels of Fig. \ref{fig:n1-galz} for the 3.6$\mic$ and 24$\mic$ populations.  In each panel, the LFs for each sub-class are shown broken down into each redshift bin.  From this procedure we see that the 3.6$\mic$ late-type and starburst galaxy populations increase in activity out to z$\sim$1.  The early-types do not contribute significantly to the evolution, and are consistent with a population formed at high redshift and then dimming as they age.  The evolution in the 24$\mic$ galaxies is also seen to be due to late-types and starbursts in the range z$\sim0.25$ to 1 and at both wavelengths the results would support an increase in the frequency of starbursts (and hence by implication, mergers) in the past.

  The study of \cite{Fransechini2006} combined multi-wavelength data in the Chandra Deep Field South, including very deep optical and near-IR photometry, deep (S(3.6$\mic)>$1 $\mu$Jy) IRAC data and complementary optical spectroscopy.  Importantly, they were able to differentiate reliably between early- and late-type classes out to high redshift and thus confirmed evidence for `downsizing' occurring in galaxy formation such that more massive galaxies completed their star formation and star accumulation
earlier than less massive ones.  Their interpretation was that of a morphological shift from the star-forming to the passively evolving phase (due to mergers and gas reduction) from high to low redshift, something only hinted at by the population breakdown shown in Fig. \ref{fig:n1-galz}.  It is of interest to note that the 3.6$\mic$ LF for starbursts (bottom-left panel of Fig. \ref{fig:n1-galz}) shows little evolution at high luminosity and more marked evolution at lower luminosity, as would be expected if less massive/luminous galaxies remained active to later epochs.

The depth of the survey does not allow strong conclusions to be made about the behaviour of early-types  above z$\sim$1. The elliptical template used is of order 10 Gyr in age, though stellar synthesis models indicate that this should be reasonably similar to galaxies as young as 4 Gyr after an instantaneous burst (e.g. \citealt{Bruzual1993ApJ...405..538B}; \citealt{Bruzual2003MNRAS.344.1000B}).  Galaxies younger than this will appear bluer and more like intermediate-type galaxies.  Since the age of the Universe ranges from $\sim3.1$ Gyr to $\sim5.6$ Gyr in the range 2$<$z$<$1 we would expect few galaxies to be fit well by our elliptical template at these redshifts.
\subsection{Type 1 AGN luminosity functions with redshift}
\label{subsec:lf_qso_lf}
\begin{figure}
\begin{center}
\includegraphics[angle=90]{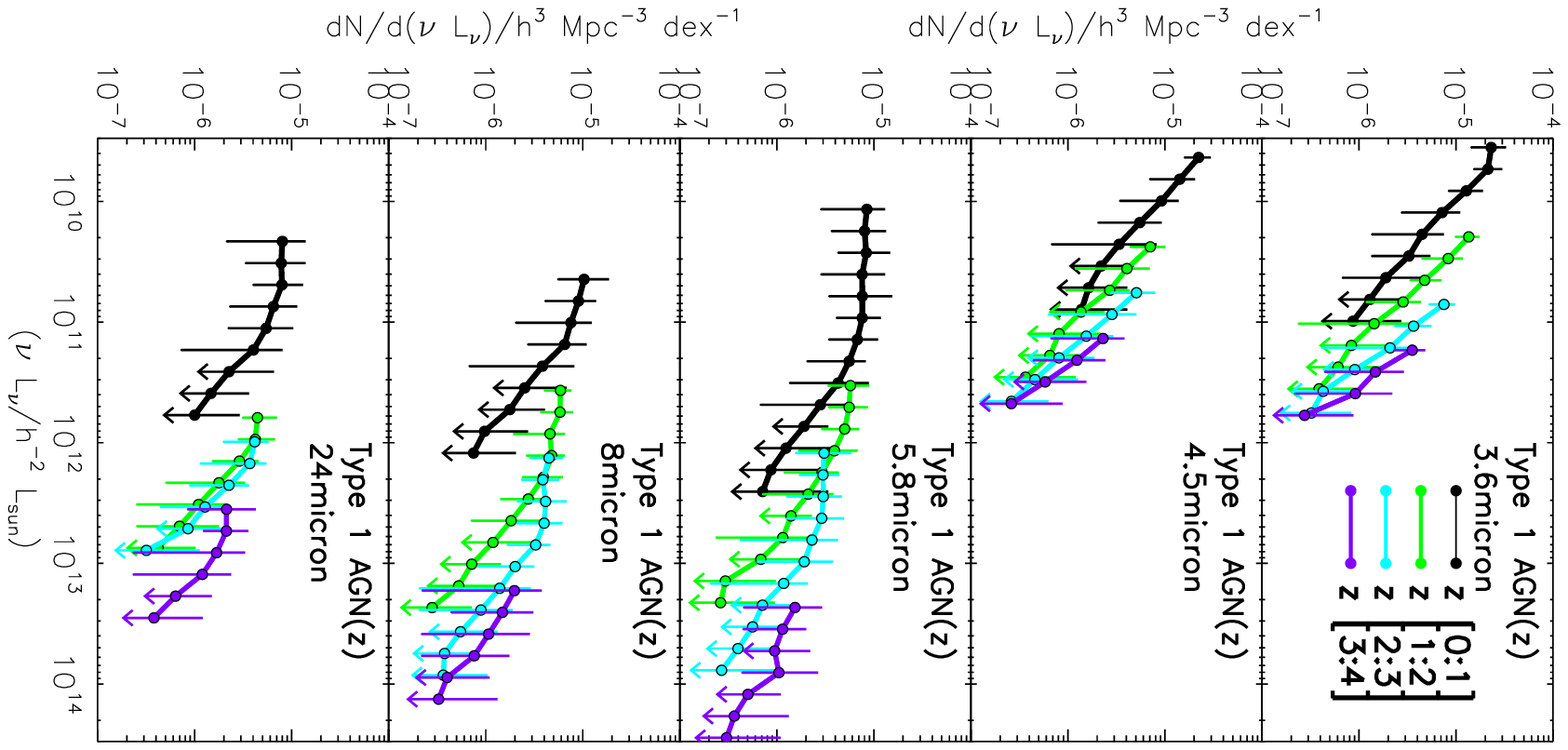}
\caption{\scriptsize{{\bf Type 1 AGN luminosity functions.} Luminosity functions for sources fit by I{\scriptsize MP}Z as Type 1 AGN, split into redshift bins of [0 to 1] (black); [1 to 2] (green); [2 to 3] (cyan); [3 to 4] (purple).}
\label{fig:n1-agn}}
\end{center}
\end{figure}

 The Type 1 AGN LFs are plotted as a function of redshift in each $Spitzer$ band in the Fig. \ref{fig:n1-agn}.  Since the Type 1 AGN are a smaller population and suffer less precise redshifts, the resulting LFs are noisier.  However, the results are clear enough to show positive evolution out to at least the third redshift bin (2$<$z$<$3), with the strongest evolution in the range 1$<$z$<$2.  

One feature is seen at 24$\mic$ where the LF in the 1$<$z$<$2 and 2$<$z$<$3 is very similar (as compared to 8$\mic$ where evolution is seen when moving from 1$<$z$<$2 to 2$<$z$<$3).  The statistical significance of this would need large-scale spectroscopic analysis for a proper interpretation but is likely to be, at least in part, due to the SED shape in these ranges - although both 8 and 24$\mic$ are being dominated by hot dust emission from an AGN for these redshifts, the 8$\mic$ emission corresponds to the rising part of the torus just after the minimum at 1$\mic$ whilst 24$\mic$ traces the peak of the torus emission, plus silicate emission.  Thus variations in the exact SED could lead to differences in the observed LF in these bands with redshift, with, for example, the presence of the silicate feature in the 24$\mic$ band at z$\sim$1.3 boosting the 1$<$z$<$2 LF luminosities.
\section{Other studies}
\label{sec:lfstudies}
\subsection{Galaxies}
\label{sec:lfstudies_gal}
\begin{figure}
\begin{center}
\includegraphics[width=8.5cm]{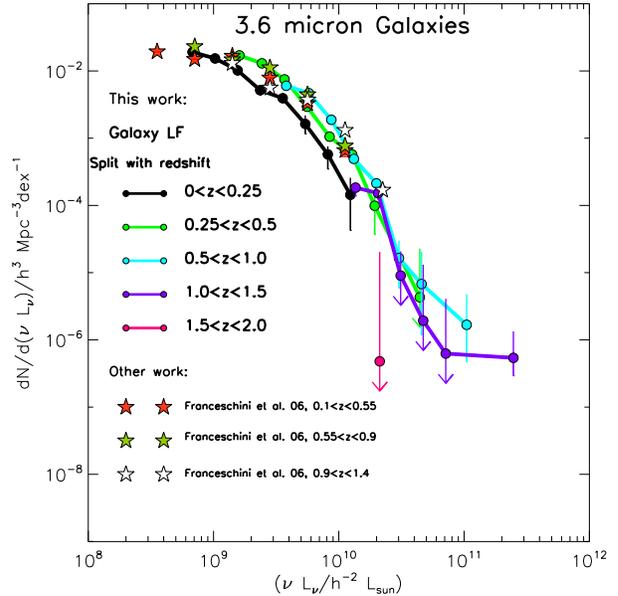}
\caption{\scriptsize{{\bf Comparison of 3.6$\mic$ galaxy LF to similar work}.  Solid circles joined by solid lines is the 3.6$\mic$ galaxy LF of this work, split into redshift bins of [0 to 0.25] (black); [0.25 to 0.5] (green); [0.5 to 1] (cyan); [1 to 1.5] (purple); [1.5 to 2] (pink).  The results of Franceschini et al. (2006) are shown as stars, for the redshift intervals [0.1 to 0.55] (red); [0.55 to 0.9] (green); [0.9 to 1.4] (white).}
\label{fig:36gal_comp}}
\end{center}
\end{figure}

The IRAC 3.6 and 4.5$\mic$ galaxy luminosity functions can be compared to near-IR luminosity functions in the J (1.25$\mic$) and K$_s$ (2.2$\mic$) bands which, similarly, probe the stellar mass in galaxies.
  The J and K$_s$ luminosity functions of \cite{Pozzetti2003AA...402..837P} found no evidence of steepening of the faint end of the LFs up to a redshift of $\sim1.3$ and no evolution of the faint end, whilst finding positive luminosity evolution at the bright end of the LFs.  Their mild evolution to z$\sim1.5$ is in agreement with previous indications (\citealt{Cowie1996AJ....112..839C}; \citealt{Cole2001MNRAS.326..255C}; \citealt{Cohen2002ApJ...567..672C}; \citealt{Feulner2003MNRAS.342..605F}).  The more recent 3.6$\mic$ LF of \cite{Fransechini2006}, for which half the sources had spectroscopic redshifts, extends out to z$\sim$1.4 and demonstrates a trend for increasing luminosity consistent with the dominant stellar populations becoming younger at higher redshift.\\
  Fig. \ref{fig:36gal_comp} compares the 3.6$\mic$ Franceschini et al. result to this work's 3.6$\mic$ galaxy LF.  The agreement between the two works is excellent, and can be considered a validation of the photometric redshift and mid-IR fitting approach used here.  Though not over-plotted for reasons of clarity the Pozzetti et al. results, shifted from  K$_s$-band to 3.6$\mic$, are also in close agreement.\\ 
  
\begin{figure}
\begin{center}
\includegraphics[width=8.5cm]{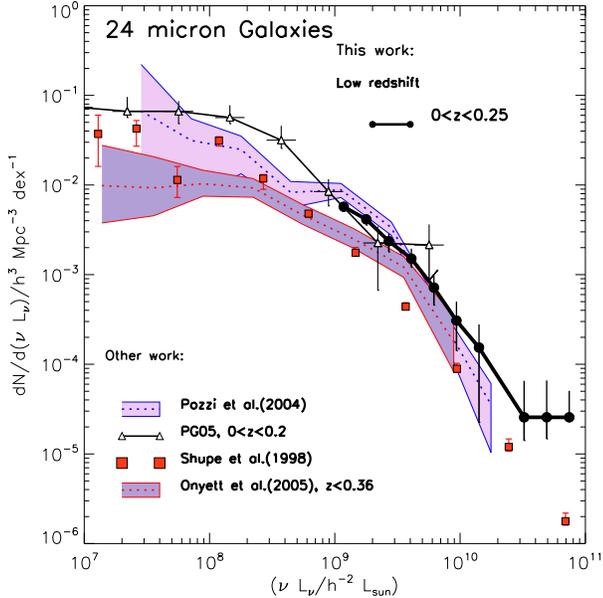}
\caption{\scriptsize{{\bf Comparison of 24$\mic$ galaxy LF to similar work: Low redshift}.  The solid coloured lines joining the circles are the LF of this work.  This is then compared to the 25$\mic$ LF of Shupe et al. (1998);  to the 15$\mic$ LFs of Pozzi et al. (2004); to the spectroscopic 24$\mic$ LF of Onyett et al. (2005) from the SWIRE Lockman Vaidation Field; and to the 24$\mic$ LF of PG05.  The Onyett et al. spectroscopic LF is not corrected for completeness in the faint bins.   For the purposes of comparison all points have been converted to the same cosmology and shifted to 24$\mic$ (when necessary) using an assumed continuum shape of -0.5.  The reader is directed to the plot for the key.
}
\label{fig:24g_compr1}}
\end{center}
\end{figure}

\begin{figure}
\begin{center}
\includegraphics[width=8.5cm]{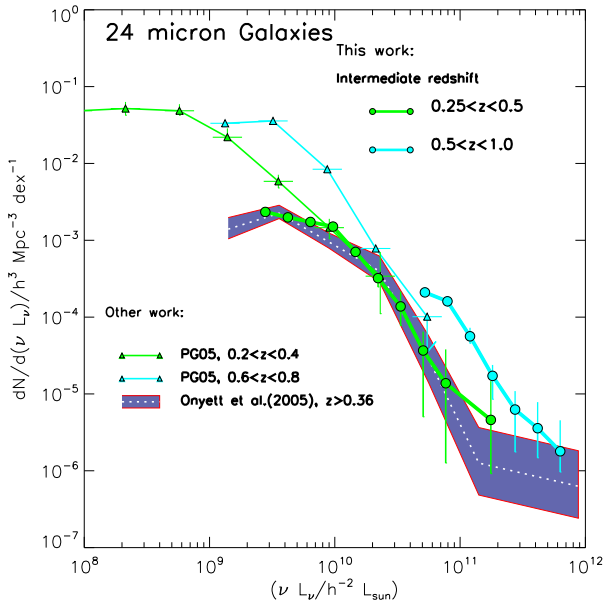}
\caption{\scriptsize{{\bf Comparison of 24$\mic$ galaxy LF to similar work: Intermediate redshift}.  Similar to Fig. \ref{fig:24g_compr1}, now for intermediate redshifts.
}
\label{fig:24g_compr2}}
\end{center}
\end{figure}

\begin{figure}
\begin{center}
\includegraphics[width=8.5cm]{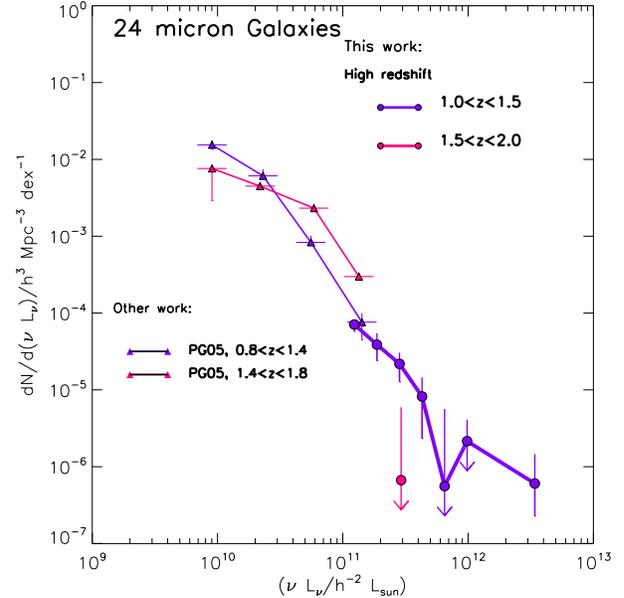}
\caption{\scriptsize{{\bf Comparison of 24$\mic$ galaxy LF to similar work: High redshift}.  Similar to Fig. \ref{fig:24g_compr1}, now for high redshifts.
}
\label{fig:24g_compr3}}
\end{center}
\end{figure}

The 24$\mic$ galaxy LF is compared to the local luminosity function at 25$\mic$ constructed by \cite{Shupe1998ApJ...501..597S}  from $IRAS$ data;  to the $ISO$ 15$\mic$ LF of \cite{Pozzi2004ApJ...609..122P}; to the deep 24$\mic$ LF of PG05 and to the spectroscopic 24$\mic$ LF of \cite{Onyett05} from the SWIRE Lockman Validation Field.  This comparison is split into three broad redshift ranges and LFs have, where necessary, been converted to 24$\mic$ using an assumed continuum shape of -0.5 for comparison purposes:\\ 
In Fig. \ref{fig:24g_compr1} the `low redshift', z$<$0.25, 24$\mic$ galaxy LF is considered. Shupe at el.'s local 25$\mic$ LF follows the low-redshift sample of our 24$\mic$ LF well, as does the 15$\mic$ LF of Pozzi et al. (median redshift of 0.18) and the low-redshift sample from Onyet et al. (z$<$0.36).  The comparison to PG05 (z$<$0.2) demonstrates the complementary nature of the datasets used - the deeper PG05 data is able to probe the LFs to fainter luminosities whilst the wider SWIRE data allows one to go to higher luminosities.  In the region of overlap the two results agree well.  The bright upturn seen in our z$<$0.25, 24$\mic$ galaxy LF for L$>$3E10 $\nu$L$_{\nu}$/h$^{-2}$L$_\odot$ can be plausibly attributed to the influence of obscured Type 2 AGN which remain in our `galaxy' selection and start to have an influence at the highest luminosity (particularly at these low redshifts where Type 2 AGN are more numerous).\\
In Fig. \ref{fig:24g_compr2} the `intermediate redshift', 0.25$<$z$<$1, 24$\mic$ galaxy LFs are 
compared to the higher-redshift sample of Onyett et al. (z$>$0.36), which exhibits a similar strong evolution at the bright-end to the intermediate-redshift bins of this work.   The median redshift of the Onyett et al. sample is 0.29, with 38\% of the sample in the z$>$0.36 bin and the most distant source at z$\sim$2.9.  The intermediate redshift LFs from PG05 are also plotted.\\  
PG05's highest redshift results;  0.8$<$z$<$1.8, are compared to ours; 1.0$<$z$<$2.0, in Fig. \ref{fig:24g_compr3}.  Despite the slightly disparate redshift bin ranges and luminosities sampled, the two results are surprisingly good for these distant populations.

In summary, then, the SWIRE LFs derived using a large photometric dataset have shown good agreement with other comparitive works, whilst, crucially, extending the LFs to high luminosity at high redshift where the results show continued evolution.

\begin{center}{\it Comparison to GALFORM+GRASIL}\end{center}
Finally, we compare our 3.6$\mic$ and 24$\mic$ galaxy LFs to model
predictions from (\citealt{Lacey2005}; Lacey et al. 2006, in prep.) that
combine output from a semi-analytic galaxy formation model (GALFORM,
\citealt{Baugh2005}; \citealt{Cole2000}), with the GRASIL
spectrophotometric code (\citealt{Silva1998ApJ...509..103S};
\citealt{Granato2000ApJ...542..710G}).  This GALFORM+GRASIL model
firsts calculates the mass assembly and star formation histories of
galaxies in the framework of the $\Lambda$CDM cosmology, and then
computes self-consistent luminosities for each galaxy, from the far-UV
to the sub-mm, from which LFs can be readily calculated.  

\begin{figure}
\begin{center}
\includegraphics[width=8.5cm]{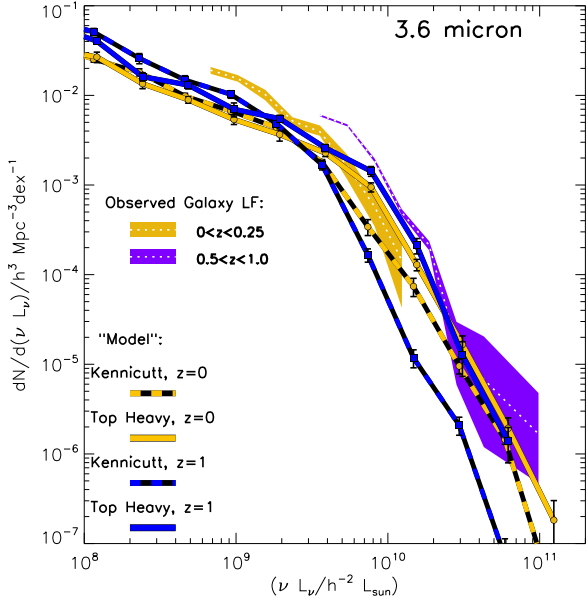}
\caption{\scriptsize{{\bf Comparison of 3.6$\mic$ galaxy LF to model}.  The shaded regions are the observed galaxy LF at 0$<$z$<$0.25 (orange) and 0.5$<$z$<$1 (lilac), with the dotted line marking the LF and the region representing the 1$\sigma$ errors.  Overplotted are the models of Lacey et al. for the redshifts 0 (orange) and 1 (blue), split into Kennicutt (1983) and Top-Heavy IMF versions as striped and solid lines respectively.
}
\label{fig:36gal_compmod}}
\end{center}
\end{figure}

\begin{figure}
\begin{center}
\includegraphics[width=8.5cm]{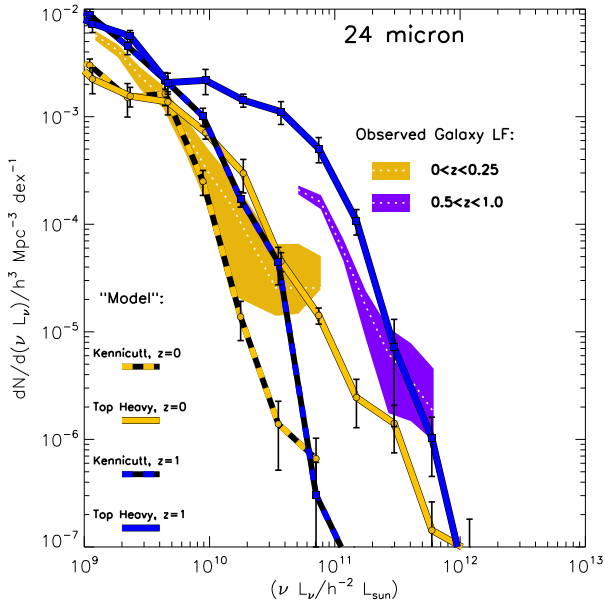}
\caption{\scriptsize{{\bf Comparison of 24$\mic$ galaxy LF to model}.  The shaded regions are the observed galaxy LF at 0$<$z$<$0.25 (orange) and 0.5$<$z$<$1 (lilac), with the dotted line marking the LF and the region representing the 1$\sigma$ errors.  Overplotted are the models of Lacey et al. for the redshifts 0 (orange) and 1 (blue), split into Kennicutt (1983) and Top-Heavy IMF versions as striped and solid lines respectively.}
\label{fig:24g_compmod}}
\end{center}
\end{figure}

In Fig. \ref{fig:36gal_compmod} and Fig. \ref{fig:24g_compmod}, we
compare our 3.6$\mic$ and 24$\mic$ galaxy LFs to two models which
differ only in the IMF assumed. In one model, labelled ``Kennicutt''
in the figures, all stars form with a solar neighbourhood IMF (we use
the \citealt{Kennicutt1983ApJ...272...54K} parametrization, which is
close to Salpeter above $1 M_{\odot}$, but with a flatter slope below
this), while in the other model, labelled ``top-heavy'', stars formed
in bursts triggered by galaxy mergers are assumed to have a top-heavy
IMF, with a much larger fraction of high-mass stars than in the solar
neighbourhood. The parameters of the ``top-heavy'' model are identical
to those used in \citep{Baugh2005}, who found that a top-heavy IMF was
essential for their CDM-based model to reproduce the observed numbers
of faint sub-mm galaxies, and have not been adjusted to match the data
from SPITZER.  It can be seen that the IMF models diverge at high luminosity and redshift, where the top-heavy IMF produces more sources of high luminosity.  This is most apparent at 24$\mic$ where more UV photons from the most massive stars are re-radiated by dust but even at 3.6$\mic$ there is a clear increase for the top-heavy IMF since more high-mass stars evolve though the Asymptotic Giant phase.  We see that at both 3.6$\mic$ and 24$\mic$, our measured
LFs are consistent with the model LFs based on a top-heavy IMF, and
lie some distance from the predictions for a standard IMF, in
particular under-estimating the number of luminous galaxies,
especially at higher redshift.

This is a very interesting result. While our measurements of the LF
cannot uniquely determine the form of the IMF, it does seem that, when
combined with galaxy formation models, the observed evolution in the
LF at both 3.6$\mic$ and 24$\mic$ favours an IMF in starbursts which
is skewed towards higher masses compared to the solar
neighbourhood. Such a top-heavy IMF is also suggested by observations
of chemical abundances in elliptical galaxies and the intra-cluster
gas in galaxy clusters, which seem to require larger yields of heavy
elements than are predicted for a standard IMF
(\citealt{Nagashima2005a}; \citealt{Nagashima2005b}).  A top-heavy IMF
would also help significantly in reconciling theroretical models of
galaxy formation based on CDM with results from extragalactic sub-mm
surveys.  The large numbers of distant IR- luminous sources found by
SCUBA (\citealt{Eales1999ApJ...515..518E}; \citealt{Borys2003MNRAS.344..385B}; \citealt{Scott2002MNRAS.331..817S}) pose a significant
challenge to CDM-based models, and early versions struggled to explain
both the observed sub-mm counts, and the inferred star formation rates
in these sources (e.g. \citealt{Granato2000ApJ...542..710G}).  Later
models have therefore invoked a variety of solutions to this problem,
including a top-heavy stellar IMF, as proposed by \citep{Baugh2005};
in this way the same sub- mm flux can be generated with substantially
lower star formation rates. Our results lend further observational
support to this idea which, if true, means that the colossal star
formation rates (up to several thousand M$_\odot$ per year) currently
inferred for distant sub-mm sources may need to be revised
significantly downwards, possibly to levels comparable to those seen
in local ULIRGs (e.g. \citealt {Farrah2003MNRAS.343..585F}).
\subsection{Quasars}
\label{subsec:lfstudies_qso}
The evolution of quasars is well established.  Most studies agree that the space density of quasars above a given luminosity increases by a large factor ($\sim$ 50) out to z$\sim2$, then levels out, and then decreases to z$\sim$5 (e.g. \citealt{Richards2006astro-ph}; \citealt{Croom2004MNRAS.349.1397C}).  The decrease from redshift z$\sim2$ to the present day can be interpreted as due to; a reduction in the merger rate; a decrease in the cold gas available for fueling; and an increase in the timescale of gas accretion \citep{Kauffmann2000MNRAS.311..576K}.

  In this work we see a strong increase in the space density at a given luminosity from the 0$<$z$<$1 to the 1$<$z$<$2 bin for all bands and further increase in the 2$<$z$<$3 bin.  At higher redshift the number statistics become poorer but are consistent with a constant/rising space density.  The recent 8$\mic$ type 1 quasar evolution of \cite{Brown2006ApJ...638...88B}, derived from S(24$\mic$)$>$1mJy sources with 1$<$z$<$5, found the peak quasar space density to be at z=2.6$\pm{0.3}$ whilst here the 8$\mic$ luminosity evolution shown in Fig. \ref{fig:agn_ple} rises strongly to 2$<$z$<$3 and then flattens off.

  Hence our broad-brush photometric redshift approach finds a similar result to other quasar studies whilst extending the Type 1 AGN LFs to the $Spitzer$ wavebands.   However, due to the depth of the SWIRE survey we are unable to fully observe the onset of strong decline at the highest redshifts.  From optical studies this strong decline in Type 1 AGN activity occurs above redshifts of $\sim$3--4, which is our highest redshift bin, and has been seen in numerous quasar studies since the early `80's (e.g \citealt{Osmer1982ApJ...253...28O}; \citealt{Warren1994ApJ...421..412W}; \citealt{Schmidt1995AJ....110...68S}; \citealt{Kennefick1995AJ....110.2553K}; \citealt{Fan2001AJ....121...54F}).  This dramatic drop can be interpreted as marking the epoch of peak formation of quasars, at around ten per cent of the current age of the Universe.  As such, it has strong implications for models of black hole formation, fuelling and evolution (e.g. \citealt{Haehnelt1993MNRAS.263..168H}; \citealt{Granato2001MNRAS.324..757G}; \citealt{Hatziminaoglou2003MNRAS.343..692H}; \citealt{Borys2005ApJ...635..853B}).

\begin{figure}
\begin{center}
\includegraphics[width=8cm]{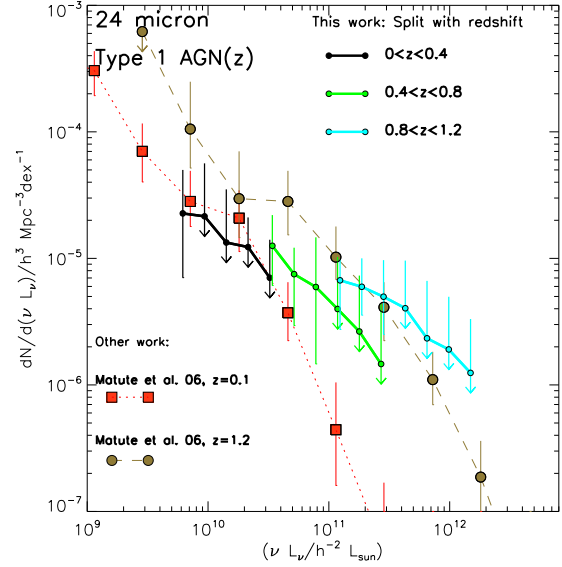}
\caption{\scriptsize{{\bf Comparison of 24$\mic$ Type 1 AGN LF to similar works}.  Comparison to the 15$\mic$ ELAIS AGN LF of Matute et al. (2006).  For the purposes of comparison all points have been converted to the same cosmology and shifted to 24$\mic$ using an assumed continuum shape of -0.5.
}
\label{fig:24agn_comp}}
\end{center}
\end{figure}

In Fig. \ref{fig:24agn_comp} the 24$\mic$ Type 1 AGN LF is compared to the 15$\mic$ ELAIS AGN LF of \cite{Matute2006} (consistent with the earlier LF of \citealt{Matute2002MNRAS.332L..11M}).   
The Matute et al. LF is split into a z=0.1 population and a z=1.2 population, which trace a similar behaviour to the evolution with redshift found in this work.  The broad redshift bins of this work make a direct comparison problematic since wider redshift ranges have the effect of flattening the resulting LF, however the results are consistent.  For closer comparison, the LF from this work has been re-binned into narrower redshift intervals, though the resulting low numbers of 24$\mic$ Type 1 AGN in the bins does result in upper limits, particularly at higher luminosity and for the lowest redshift bin (z$<$0.4).
\section{Luminosity evolution, energy and SFR density}
\label{sec:lf-fits}
\subsection{Schecter and double power-law fits}
\label{subsec:lf-schecpfits}
The LFs calculated here demonstrate the evolution of both galaxies and Type 1 AGN with redshift.  We now model these LFs using Schecter Functions (\citealt{Schechter1976ApJ...203..297S}) and double power-law fits.  The Schecter parameterisation is often applied to UV/optical galaxy LFs, whilst double power-law fits have been found to be a good fit to the local infrared population (e.g. \citealt{Lawrence1986MNRAS.219..687L}; \citealt{Saunders1990MNRAS.242..318S}; \citealt{Takeuchi2003ApJ...587L..89T}).  
 This can be broadly explained since star formation acts as a power-law in the luminosity function and stellar mass acts as an exponential drop-off - a luminosity function can then be seen as a convolution of the two regimes, weighted by the relative dominance of each (determined by the rest-frame wavelength).

The Type 1 AGN LF is often modelled by a double power-law \citep{Boyle1988MNRAS.235..935B}.  For example, \cite{Pei1995ApJ...438..623P} used a double power-law model where the bright and faint ends were independent of redshift, with the break occurring at a luminosity that was dependent on redshift.

The Schecter parameterisation is as follows:
\begin{equation}
 \label{eqn:lumfunc}
 \Phi(L)dL/L^{*} = \Phi^{*}\bigg(\frac {L}{L^{*}}\bigg)^{\alpha}e^{-L/L^{*}}dL/L^{*},
\end{equation}

whilst the double power-law is described by:
\begin{equation}
 \label{eqn:lumfunc2p}
 \Phi(L)dL/L^{*} = \Phi^{*}\bigg(\frac {L}{L^{*}}\bigg)^{1-\alpha}dL/L^{*},  L<L^{*}
\end{equation}
and
\begin{equation}
 \label{eqn:lumfunc2p2}
 \Phi(L)dL/L^{*} = \Phi^{*}\bigg(\frac {L}{L^{*}}\bigg)^{1-\beta}dL/L^{*},  L>L^{*}
\end{equation}

Our approach to fitting the LFs is determined by the wavelength under consideration and the extent to which the data can break parameter degeneracies.  Thus, we choose to split the analyses in the following manner:
\subsection{5.8 and 8$\mic$ galaxies, and Type 1 AGN LFs - luminosity evolution}
\label{subsubsec:fixpfits}
\begin{figure}
\begin{center}
\includegraphics[width=8cm]{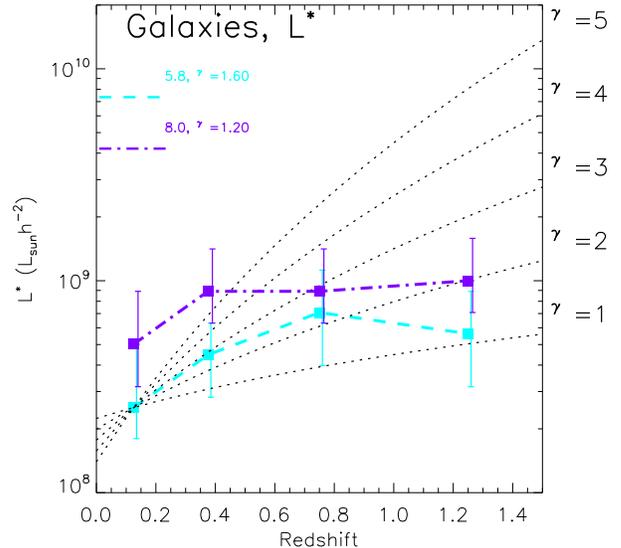}
\caption{\scriptsize{{\bf L$^*$ (L$_\odot$h$^{-2}$) for 5.8 and 8$\mic$ galaxies.}  5.8 and 8$\mic$ parameters result from PLE power-law fits.  The overlaid dotted lines, renormalised to the 5.8$\mic$ value in the first redshift bin, show the trend if L$^*\Phi^*\propto (1+z)^\gamma$, for various $\gamma$.  The best-fit $\gamma$ are $\gamma$=1.6$^{+0.7}_{-0.6}$ at 5.8$\mic$ and $\gamma$=1.2$^{+0.4}_{-0.5}$ at $8\mic$ (1$\sigma$ errors in the parameter space).}
\label{fig:gal_ple}}
\end{center}
\end{figure}

\begin{figure}
\begin{center}
\includegraphics[width=8cm]{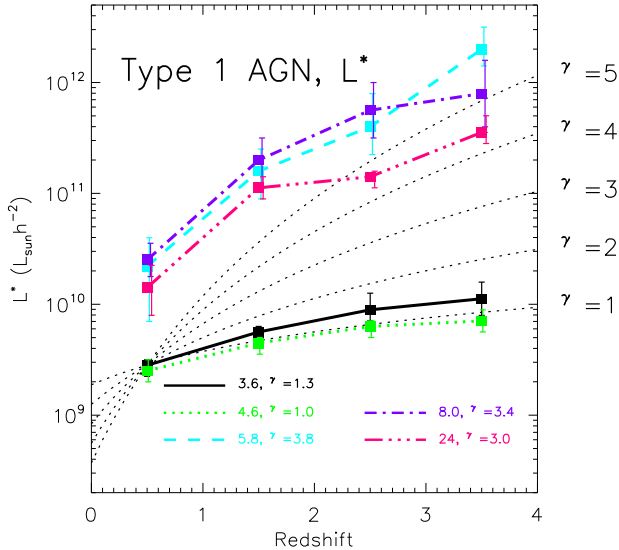}
\caption{\scriptsize{{\bf L$^*$ (L$_\odot$h$^{-2}$) for Type 1 AGN.}  L$^*$ values result from PLE power-law fits.  The overlaid dotted lines, renormalised to the 3.6$\mic$ value in the first redshift bin, show the trend if L$^*\propto$ (1+z)$^\gamma$, for various $\gamma$.  The best-fit $\gamma$ are, respectively, $\gamma$=1.3$^{+0.1}_{-0.1}$, 1.0$^{+0.1}_{-0.1}$, 3.8$^{+0.4}_{-0.3}$, 3.4$^{+0.4}_{-0.3}$, 3.0$^{+0.5}_{-0.4}$ (1$\sigma$ errors in the parameter space).}
\label{fig:agn_ple}}
\end{center}
\end{figure}

Double power-law functions are fit to the 5.8 and 8$\mic$ galaxy LFs and to the Type 1 AGN LFs (note that Schecter fits were attempted for the 5.8$\mic$ galaxy LF but these did not give good results, most obviously at the bright end).  However, the luminosity ranges sampled by these SWIRE LFs do not allow us to constrain the faint-end well (and hence the faint-end slope and the break where the bright-end slope takes over).  Thus, noting that the bright-end slopes are well described by the same slope across the redshift bins, we choose to explore the evolution of L$^*$ only, by fixing $\alpha$, $\beta$, and $\Phi^*$.   We set the faint-end slope to $\alpha=1.3$ (cf. PG05's slope of 1--1.3 for their 24$\mic$ sample; \citealt{Zheng2006ApJ...640..784Z}'s value of 1.2$\pm$0.3 from analysis of the 24$\mic$ LFs of \citealt{LeFloch052005} and the local value for 60$\mic$ IRAS galaxies of 1.23 from \citealt{Takeuchi2003ApJ...587L..89T} and 1.09 from \citealt{Saunders1990MNRAS.242..318S}).  The bright-end slope for the 5.8 and 8$\mic$ galaxy LFs are set by the slope in the lowest redshift bin (0$<$z$<$0.25) where it is best-constrained, giving $\beta$=2.5 and 2.6 respectively.  For the Type 1 AGN LFs a shallower slope of $\beta$=2.2 is able to give a good fit across all the bands.  The density parameter $\Phi^*$ is set to 5E-2 h$^3$dex$^{-1}$Mpc$^{-3}$ for the 5.8 and 8$\mic$ galaxy LFs and to 5E-5 for the Type 1 AGN.  It is important to note that although we have chosen reasonable values, for the purposes of exploring pure luminosity evolution (PLE) the exact choice of the fixed $\alpha$ and $\Phi^*$ values only effects the overall scaling of L$^*$, not its evolution as seen by SWIRE.  The resulting PLE trends are then characterised by
\begin{equation}
\label{eqn:gamma}
L^*\propto (1+z)^\gamma.
\end{equation}
Fig. \ref{fig:gal_ple} shows the evolution of L$^*$ for the 5.8 and 8$\mic$ galaxies, whilst Fig. \ref{fig:agn_ple} shows the evolution for the Type 1 AGN.  For the two galaxy LFs L$^*$ is seen to increase out to the 0.5$<$z$<$1 resdhift bin, and then flattens off at higher redshift.  The best-fit $\gamma$ is found to be $\gamma$=1.6$^{+0.7}_{-0.6}$ at 5.8$\mic$ and $\gamma$=1.2$^{+0.4}_{-0.5}$ at $8\mic$, with the majority of the evolution occurring since z$\sim$1.  In contrast the Type 1 AGN show positive evolution in all bands, out to the highest redshifts (3$<$z$<$4).  There is a noticeably higher relative luminosity across redshift for $\lambda>$5$\mic$ in comparison to $\lambda<$5$\mic$.
The best-fit $\gamma$'s are $\gamma$=1.3$^{+0.1}_{-0.1}$ at 3.6$\mic$, $\gamma$=1.0$^{+0.1}_{-0.1}$ at $4.5\mic$, $\gamma$=3.8$^{+0.4}_{-0.3}$ at 5.8$\mic$, $\gamma$=3.4$^{+0.4}_{-0.3}$ at 8$\mic$ and $\gamma$=3.0$^{+0.5}_{-0.4}$ at 24$\mic$.  The strong evolution at longer wavelengths is consistent with other studies - e.g. \cite{Matute2006} found a pure luminosity evolution model with L$\propto$(1+z)$^{\sim2.9}$ was a good fit to their 15$\mic$ ELAIS AGN LF.
\subsection{3.6 and 4.5$\mic$ galaxies - energy density}
\label{subsubsec:endense}
\begin{figure}
\begin{center}
\includegraphics[width=8cm]{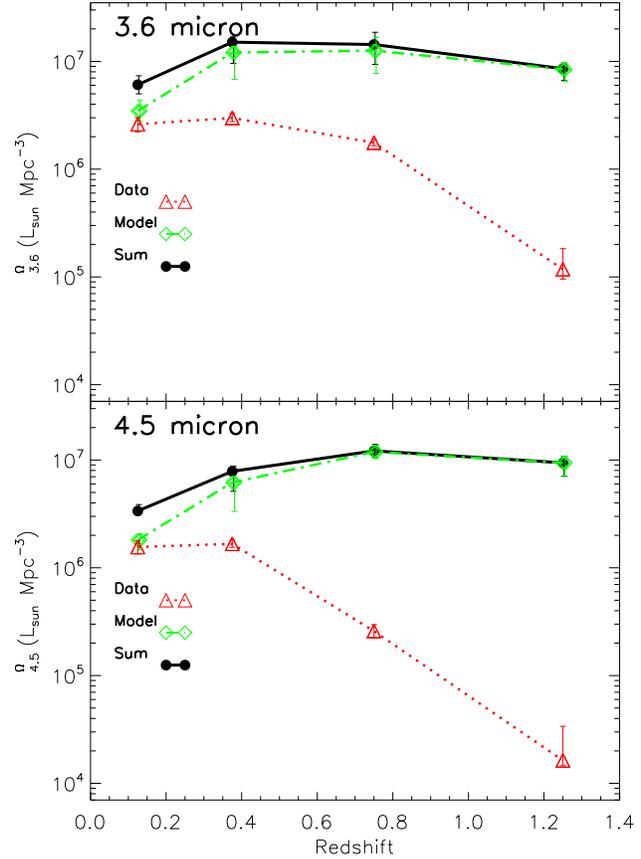}
\caption{\scriptsize{{\bf Integrated energy density $\Omega$ (L$_\odot$Mpc$^{-3}$) for 3.6 and 4.5$\mic$ galaxies.}  $\Omega_{3.6}$ and $\Omega_{4.5}$ (black circles) result from the combination of the observed data (red triangles) and the integrated Schecter-fits outside of the luminosity range where there was data (green diamonds).}
\label{fig:endense}}
\end{center}
\end{figure}

We choose to fit Schecter functions to the IRAC 3.6 and 4.5$\mic$ galaxy LFs.  As the data are sufficient to constrain the parameter space we allow $\alpha$, $\Phi^*$, and L$^*$ to vary, except for the 1$<$z$<$1.5 bin where we use the $\alpha$ value fit to the 0.5$<$z$<$1 slope due to the reduced data at this high redshift.  We are then able to integrate the resulting Schecter functions over luminosity in order to determine the total energy density, $\Omega$, provided by 3.6 and 4.5$\mic$ galaxies at each redshift.  This integral is plotted in Fig. \ref{fig:endense}, showing the contribution from the observed data, and from the Schecter-fits extrapolated to luminosities not sampled by the observed LFs.  Consider that an old passively evolving population is characterised by PLE, dimming more quickly than a no-evolution model in the K band, say (e.g. \citealt{Ellis2004MNRAS.348..165E}; \citealt{dePropris1999AJ....118..719D}).  This scenario is essentially what is found for IRAC 3.6 and 4.5$\mic$, which probe the older stellar populations (indeed a PLE Schecter fit describes the change with redshift almost as well as one in which both L$^*$ and $\Phi^*$ are allowed to vary).  Fig \ref{fig:endense} shows that for both bands, the energy density gradually increases out to z$\sim$0.5--1 but flattens, or even declines, at higher redshift.  
\subsection{24$\mic$ galaxies - energy and SFR density}
\label{subsubsec:sfh}
For the 24$\mic$ galaxies we choose to fit a double power law.  Again, the data is not able to constrain the faint-end slope and the turnover and we fix the faint-end slope to $\alpha$=1.3 and $\Phi^*$=5E-2 h$^3$dex$^{-1}$Mpc$^{-3}$ (as before in \S\ref{subsubsec:fixpfits}).  However, as shown in \S\ref{sec:lfstudies_gal} the agreement with the PG05 LF (which is at fainter luminosities) is good and this choice of $\alpha$ and $\Phi^*$ gives double power-law fits to our data that also match well with the PG05 data at the faint-end and turnover location.  This gives us confidence in extending the analysis to calculate the 24$\mic$ energy density, $\Omega_{24}$, by summing the contribution of our data and the integral of the double power-law outside of the luminosity range in which we have data (as in \S\ref{subsubsec:endense}).  For the 24$\mic$ analysis we now go further and use $\Omega_{24}$ to obtain an estimate of the total (8--1000$\mic$) IR energy density, $\Omega_{IR}$, and the resulting implied cosmic-SFR density:\\

{\it Total IR luminosity}\\
The correlation between rest-frame 12-15$\mic$ and total IR luminosity for local IR-luminous galaxies has been demonstrated numerous times (e.g. \citealt{Papovich2002ApJ...579L...1P}; \citealt{Chary2001ApJ...556..562C}) to be good, with a scatter of around 0.15 dex.  \cite{Bell2005ApJ...625...23B} calculated a conversion from observed-frame 24$\mic$ to total IR luminosity using the full range of model template spectra of \cite{Dale2001ApJ...549..215D}, with a mean correction factor of, to within 20 per cent, 
\begin{equation}
\label{eqn:totalIR}
{\rm L}_{IR}\sim 10\nu\L_{\nu} (\rm{24}\mic).
\end{equation}
The mean template spread around this mean was larger, $\sim$0.3 dex.  Similar results have been found by the study of \cite{Chary2001ApJ...556..562C} for $ISO$ and $IRAS$ galaxies, and by \cite{LeFloch052005} (see their figure 8).  Thus, we convert our $\Omega_{24}$ to $\Omega_{IR}$ using Eqn. \ref{eqn:totalIR}, which should be accurate to a factor 2 or so \citep{Bell2005ApJ...625...23B}.\\

{\it Star formation rate density}\\
We use $\Omega_{IR}$ to obtain the SFR density using the relation of \cite{Kennicutt1998ARA&A..36..189K};
\begin{equation}
\label{eqn:SFR}
{\rm SFR (M_{\odot}yr^{-1})}=1.71{\rm x}10^{-10} {\rm L}_{IR} ({\rm L_{\odot}}).
\end{equation}
As discussed in PG05, this conversion results in further uncertainty - the SFRs are subject to photometric redshift uncertainties, differences between different model templates, and the validity of extrapolating low-redshift relationships such as that of Eqn. \ref{eqn:totalIR} to higher redshift.  Thus the resulting SFRs can be expected to be subject to an additional systematic error of up to a factor 2--3.

\begin{figure}
\begin{center}
\includegraphics[width=8cm]{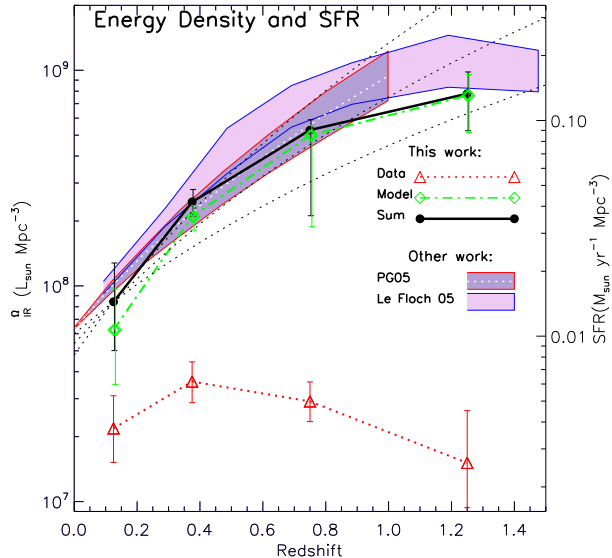}
\caption{\scriptsize{{\bf  Integrated IR energy density $\Omega_{IR}$ and SFR density from 24$\mic$ galaxies.}  $\Omega_{IR}$ (black circles) results from the combination of the observed data (red triangles) and the integrated two-power fits outside of the luminosity range where there was data (green diamonds) to obtain $\Omega_{24}$.  This is then converted to $\Omega_{IR}$ (L$_\odot$Mpc$^{-3}$) and SFR (M$_{\odot}$yr$^{-1}$Mpc$^{-3}$) as detailed in the text.  The overlaid black-dotted lines, renormalised to the value in the first redshift bin, show the best-fit trend and 1$\sigma$ spread if $\Omega_{IR}\propto$ (1+z)$^\gamma$, with best-fit $\gamma$=4.5$^{+0.7}_{-0.6}$.  For comparison the $\Omega_{IR}$ results from PG05 and Le Floc'h et al. (2005) are also plotted.
\label{fig:sfr}}}
\end{center}
\end{figure}

The cosmic IR energy density, $\Omega_{IR}$ (left axis), and SFR density (right axis) are plotted in Fig. \ref{fig:sfr}, showing the contribution from the observed data, and from the two-power fits extrapolated to luminosities not sampled by the observed 24$\mic$ LF.  For clarity the additional systematic uncertainties resulting from using Eqn. \ref{eqn:totalIR} and \ref{eqn:SFR} for conversion are not included in the plot.  In contrast to IRAC, strong positive evolution, characterised by $\Omega_{IR}\propto$ (1+z)$^\gamma$ with $\gamma$=4.5$^{+0.7}_{-0.6}$, is seen for the 24$\mic$ galaxies, with the majority of this evolution occurring since z$\sim$1.  For comparison, $\Omega_{IR}$ results from PG05 and Le Floc'h et al. (2005) are also plotted, showing the same strong evolution.  The evolution in star formation rate density PG05 measured was $\propto$(1+z)$^{4.0\pm0.2}$ out to a redshift of 0.8, consistent with the evolution found in this work, and with the evolution of $\gamma$=3.9$^{+0.4}_{-0.4}$ to z$\sim$1, found by \cite{LeFloch052005} for their 24$\mic$-selected sample.

Note that parameter fit values with redshift resulting from \S\ref{sec:lf-fits} are given in Appendix \ref{appendixA}.
\section{Discussion and conclusions}
\label{sec:disc_conc}
This paper has extended the photometric redshift method to incorporate near-IR data (sub 5$\mic$) in the solution, improving both the accuracy and reliability of the solutions.  Application of the extended photometric redshift technique to a 6.5 square degree area in the ELAIS N1 field of the SWIRE survey has shown that the majority of the optical SWIRE population can be characterised as star-forming galaxies with a mean redshift of $\langle$z$\rangle\sim0.6$, with a tail extending out to redshifts of z$\sim1.5$.  Early-type galaxies (fit as ellipticals) make up around a fifth of the population, with a mean redshift of $\langle$z$\rangle\sim0.4$ and a tail to a redshift of around unity.  The proportion of 24$\mic$ sources that are early-type is seen to be lower than that of the shorter wavelength $Spitzer$ sources, however the results suggests that the frequency of dusty early-types is higher than pre-$Spitzer$ studies have found.  Some of these sources could in fact be heavily obscured Arp220-like objects.  In addition to the galaxies, there is a population of optical Type 1 AGN ($\sim8\%$ of the total) that extends across the full redshift range, with indication of a slight rise in their distribution to redshift two to three, and a decline thereafter.

Using the photometric redshift solutions, along with selection functions designed to account for the bias in the multi-wavelength catalogue, allowed luminosity functions to be calculated in each $Spitzer$ band, for both galaxies and quasars.  The relevant K-corrections in the optical and IR were derived using the optical/near-IR template solutions of the redshift fitting procedure, and separate mid-IR template fitting solutions applied to the longer $Spitzer$ wavebands.  Importantly, this SWIRE sample has enabled a joint analysis of galaxy and Type 1 AGN LFs using the same sample, selection and method for both populations.\\

The 3.6 and 4.5$\mic$ galaxy LFs, where the LF is probing the stellar mass, showed overall evolution consistent with the passive ageing of the dominant stellar population up to z$\sim$1.5.  The comoving luminosity density was found to evolve passively, gradually increasing out to z$\sim$0.5--1 but flattening, or even declining, at higher redshift.  Analysis with templatel type does, however, offer tantilising support for `downsizing' with starbursts showing little evolution at high luminosity and more marked evolution at lower luminosity, as would be expected if less massive/luminous galaxies remained active to later epochs.

More pronounced evolution was seen at the longest wavelength; the 24$\mic$ galaxy LF, more indicative of star formation rate and/or AGN activity, undergoes strong positive evolution, with the derived IR energy density and SFR density $\propto$ (1+z)$^\gamma$ with $\gamma$=4.5$^{+0.7}_{-0.6}$ and the majority of this evolution occurring since z$\sim$1.  This strong evolution was in agreement with other equivalent work, however the SWIRE LFs were uniquely able to show that this evolution continues out to high luminosities at high redshift (where many models start to differ).

In addition to other observational work, comparisons to the predictions of a combined semi-analytic and spectrophotometric code were made.   While we cannot provide meaningful constraints on the 
exact form of the IMF preferred by our LFs, an IMF skewed towards 
higher mass star formation in bursts compared to locally was preferred.  As a result the currently inferred massive star formation rates in distant sub-mm sources may require substantial downwards revision.\\

Although the results for the optical Type 1 AGN LFs are subject to greater uncertainty due to reduced photometric redshift accuracy and smaller population size, the analysis was able to demonstrate clear positive luminosity evolution  in all bands, out to the highest redshifts (3$<$z$<$4).  Modelling as  L$^*\propto$(1+z)$^\gamma$ gave $\gamma\sim1-1.3$ for 3.6 and 4.5$\mic$ and stronger evolution at the longer wavelengths (5.8, 8 and 24$\mic$), of $\gamma$=3.8$^{+0.4}_{-0.3}$, 3.4$^{+0.4}_{-0.3}$and 3.0$^{+0.5}_{-0.4}$ respectively.
\section{Acknowledgements}
\label{sec:lfack}
This work makes use of observations made with the $Spitzer$ Space Telescope, operated by the Jet Propulsion Laboratory, CALTECH, NASA contract 1407.  
The INT WFS data is publicly available through the Isaac Newton Groups' Wide Field Camera Survey Programme. SJO was supported by a Leverhulme Research Fellowship and PPARC grant PPA/G/S/2000/00508.  TSRB and NO were supported by PPARC studentships. TSRB would like to thank Richard Savage for advice on expectation measures and Payam Davoodi for bimodality discussions.  We extend our thanks to the referee.
\bibliography{/Users/tsb1/Documents/Work/References/BibdeskBibliog}
\bibliographystyle{mn2e}
\appendix
\section{LF functional fit parameters}\label{appendixA}
\begin{table}  
\caption{{\bf Schecter fits}, for 3.6 and 4.5$\mic$ galaxy LFs.}
\label{table:param_stats1}
\begin{center}
\begin{tabular}{|c |c c c c|}
\hline
Sample&$\alpha$&L$^*$(L$_{\odot}$h$^{-2}$)&$\Phi^*$(h$^{3}$Mpc$^{-3}$dex$^{-1}$)&Redshift\\ 
\hline
3.6$\mic$&-0.9&3.6${\rm x}10^{-3}$&5.6$^{7.1}_{4.5}{\rm x}10^{9}$&0--0.25\\ 
&-1.0&4.5${\rm x}10^{-3}$&7.9$^{8.9}_{7.1}{\rm x}10^{9}$&0.25--0.5\\
&-0.9&7.0${\rm x}10^{-3}$&6.8$^{9.1}_{7.9}{\rm x}10^{9}$&0.5--1.0 \\
&-0.9&4.0${\rm x}10^{-3}$&7.1$^{7.9}_{5.6}{\rm x}10^{9}$&1.0--1.5 \\
\hline
4.5$\mic$&-0.9&4.5${\rm x}10^{-3}$&2.8$^{3.2}_{2.2}{\rm x}10^{9}$&0--0.25\\ 
&-1.0&6.7${\rm x}10^{-3}$&3.2$^{3.5}_{2.8}{\rm x}10^{9}$&0.25--0.5\\
&-1.1&7.7${\rm x}10^{-3}$&3.2$^{3.5}_{2.8}{\rm x}10^{9}$&0.5--1.0 \\
&-1.1&6.9${\rm x}10^{-3}$&3.2$^{3.5}_{2.5}{\rm x}10^{9}$&1.0--1.5 \\
\hline
\end{tabular}
\end{center}
\end{table}
\begin{table}  
\caption{{\bf L$^*$ fit parameters}, for 5.8 and 8$\mic$ galaxy LFs.  Double power-law, fixed parameters are: $\alpha=1.3$, $\beta$=2.5 (5.8$\mic$) and 2.6 (8$\mic$), $\Phi^*$=5E-2 h$^3$dex$^{-1}$Mpc$^{-3}$. Best-fit $\gamma$ is given for L$^*\propto$ (1+z)$^\gamma$. 
}\label{table:param_stats2}
\begin{center}
\begin{tabular}{|c |c c c|}
\hline
Sample&L$^*$(L$_{\odot}$h$^{-2}$)&Redshift&$\gamma$\\ 
\hline
5.8$\mic$&2.5$^{4.0}_{2.2}{\rm x}10^{8}$ &0--0.25&1.6$^{+0.7}_{-0.6}$\\  
&4.5$^{6.3}_{2.8}{\rm x}10^{8}$ &0.25--0.5& \\
&7.1$^{11.2}_{4.0}{\rm x}10^{8}$&0.5--1.0&\\
&5.6$^{8.9}_{3.2}{\rm x}10^{8}$ &1.0--1.5&\\ 
\hline
8.0$\mic$&5.0$^{8.9}_{3.2}{\rm x}10^{8}$ &0--0.25&1.2$^{+0.4}_{-0.5}$\\  
&8.9$^{14.1}_{6.3}{\rm x}10^{8}$ &0.25--0.5& \\
&8.9$^{14.1}_{6.3}{\rm x}10^{8}$&0.5--1.0&\\
&1.0$^{1.6}_{0.7}{\rm x}10^{9}$ &1.0--1.5&\\ 
\hline
\end{tabular}
\end{center}
\end{table}
\begin{table}  
\caption{{\bf L$^*$ fit parameters}, for Type 1 AGN LFs. Double power-law, fixed parameters are: $\alpha=1.3$, $\beta$=2.2, $\Phi^*$=5E-5 h$^3$dex$^{-1}$Mpc$^{-3}$. Best-fit $\gamma$ is given for L$^*\propto$ (1+z)$^\gamma$.
}\label{table:param_stats4}
\begin{center}
\begin{tabular}{|c |c c c|}
\hline
Sample&L$^*$(L$_{\odot}$h$^{-2}$)&Redshift&$\gamma$\\ 
\hline
3.6$\mic$& 2.8$^{3.2}_{2.2}{\rm x}10^9$&0--1&1.3$^{+0.1}_{-0.1}$\\  
&5.6$^{6.3}_{5.0}{\rm x}10^9$&1--2 &\\
&8.9$^{12.6}_{7.1}{\rm x}10^9$&2--3& \\
& 1.1$^{1.6}_{0.8}{\rm x}10^{10}$&3--4&\\ 
\hline
4.5$\mic$&2.5$^{3.2}_{2.0}{\rm x}10^9$&0--1&1.0$^{+0.1}_{-0.1}$\\  
&4.5$^{5.0}_{3.6}{\rm x}10^9$&1--2 &\\
&6.3$^{7.1}_{5.0}{\rm x}10^9$&2--3 &\\
&7.1$^{8.9}_{5.6}{\rm x}10^9$&3--4&\\ 
\hline
5.8$\mic$&2.2$^{4.0}_{0.7}{\rm x}10^{10}$&0--1&3.8$^{+0.4}_{-0.3}$\\  
&1.6$^{2.5}_{0.9}{\rm x}10^{11}$ &1--2 &\\
&4.0$^{7.9}_{2.2}{\rm x}10^{11}$ &2--3 &\\
&2.0$^{3.2}_{1.4}{\rm x}10^{12}$&3--4&\\ 
\hline
8.0$\mic$&2.5$^{3.6}_{1.8}{\rm x}10^{11}$ &0--1&3.4$^{+0.4}_{-0.3}$\\  
&2.0$^{3.2}_{1.1}{\rm x}10^{11}$ &1--2& \\
&5.6$^{10.0}_{3.2}{\rm x}10^{11}$&2--3 &\\
&7.9$^{1.6}_{4.0}{\rm x}10^{11}$ &3--4&\\ 
\hline
24$\mic$&1.4$^{2.2}_{0.8}{\rm x}10^{10}$&0--1&3.0$^{+0.5}_{-0.4}$\\  
&1.1$^{1.4}_{0.9}{\rm x}10^{11}$ &1--2& \\
&1.4$^{1.6}_{1.1}{\rm x}10^{11}$ &2--3 &\\
&3.6$^{5.0}_{2.8}{\rm x}10^{11}$ &3--4&\\ 
\hline
\end{tabular}
\end{center}
\end{table}
\begin{table}  
\caption{{\bf Double power-law functional fit parameters}, for the 24$\mic$ galaxy LF with $\alpha=1.3$ and $\Phi^*$=5E-2 h$^3$dex$^{-1}$Mpc$^{-3}$. Best-fit $\gamma$ is given for L$^*\propto$ (1+z)$^\gamma$.}
\label{table:param_stats4}
\begin{center}
\begin{tabular}{|c |c c c c|}
\hline
Sample&$\beta$&L$^*$(L$_{\odot}$h$^{-2}$)&Redshift&$\gamma$\\
\hline
24$\mic$&-2.5&3.6$^{6.3}_{1.8}{\rm x}10^{8}$&0--0.25&1.6$^{+0.7}_{-0.6}$\\  
&-3.0&1.6$^{1.9}_{1.3}{\rm x}10^{9}$&0.25--0.5&\\
&-3.0&3.2$^{3.6}_{1.1}{\rm x}10^{9}$&0.5--1.0& \\
&-3.0&4.5$^{5.6}_{3.0}{\rm x}10^{9}$&1.0--1.5& \\
\hline
\end{tabular}
\end{center}
\end{table}
\label{lastpage}
\end{document}